\documentclass[prc,aps,a4paper,groupedaddress,superscriptaddress,nofootinbib,showpacs
,preprintnumbers,twocolumn]{revtex4}
\usepackage{graphicx}
\usepackage{amsfonts}
\usepackage{amssymb}
\usepackage{amsmath}
\usepackage{natbib}
\usepackage{dcolumn}
\usepackage{bm} 
\newcommand{\bwt}{\begin{widetext}}
\newcommand{\ewt}{\end{widetext}}
\newcommand{\beq}{\begin{equation}}
\newcommand{\eeq}{\end{equation}}
\newcommand{\bea}{\begin{eqnarray}}
\newcommand{\eea}{\end{eqnarray}}
\begin{document}
\title{Effect of strong magnetic fields on the pasta phase structure}
\author{R. C. R. de Lima}
\affiliation{Depto de Matem\'{a}tica, CCT, Universidade do Estado de Santa Catarina, Joinville, SC, 89219-710, Brazil}

\author{S. S. Avancini}
\affiliation{Depto de F\'{i}sica, CFM, Universidade Federal de Santa Catarina, Florian\'{o}polis, SC, CP476, 88040-900, Brazil} 

\author{C. Provid\^{e}ncia}
\affiliation{Centro de F\'{i}sica Te\'{o}rica, Dep. de F\'{i}sica, Universidade de Coimbra, 3004-516, Coimbra, Portugal} 
\date{\today}
\begin{abstract}
The effect of strong magnetic fields on the properties of the pasta structures
is calculated within a Thomas Fermi approach using relativistic mean field
models to modulate stellar matter. It is shown how quantities such as the size of
the clusters and Wigner-Seitz cells, the surface tension and the
transition between configurations are affected. It is expected that
these effects may give rise to large  stresses in the pasta phase if  the
local magnetic field suffers  fluctuations.
\end{abstract}
\maketitle

\section{Introduction}

In the bottom of the inner crust of  proto-neutron  and neutron stars, where the transition to the homogeneous core matter
occurs,  it is expected
the existence of a special matter known as
{\it pasta phase}. This phase is  a frustrated system that arises in the
competition between the strong and the electromagnetic interactions
\cite{pethick,hashimoto84,hor05,maruyama05,watanabe08,grill2012}. 
 The basic shapes of these structures were named according to their
geometry, droplets (bubbles), rods (tubes) and slabs for three, two and one dimensions,
respectively  \cite{pethick}, and the ground-state configuration is the one that minimizes the
free energy. 

The pasta phase has been studied within the Thomas Fermi approximation
 at zero and finite temperature within
different parametrizations of the relativistic non-linear Walecka model \cite{bb}
 and of the density-dependent hadronic model \cite{pasta1,pasta2,pasta3}.

It is known that magnetars, neutron stars with very strong magnetic fields of the order of
$10^{14}-10^{15}$ G at the surface, are  sources of very energetic electromagnetic 
radiation, mainly  gamma and X rays
\cite{duncan,usov,pacz}. Presently, more than 20 of 
these objects have been detected, most of them as
soft gamma repeaters (SGRs) and anomalous X-ray pulsars (AXPs)
\cite{index}. It is not clear how strong is the magnetic field in the
interior but several studies seem to indicate that 
fields stronger than
$~10^{18}$ G are not allowed.  According to the scalar
virial theorem~\cite{virial} the interior magnetic field strength could be as large as
$B \sim 1-3\times 10^{18}$ G.  Similar values were obtained in
\cite{broderick02} from  general relativistic magneto-hydrostatic
calculations, or in \cite{sedrakian13} were the vanishing of the
pressure parallel to the field  restricts  homogeneously distributed
fields to intensity below 10$^{19}$ G.

In  the present study we investigate the effect of the magnetic field
on the pasta structure. In \cite{chakrabarty97} a simple expression,
dependent on two parameters and the magnetic field intensity at the
surface, was proposed to modulate the magnetic field with
density. Taking  this expression as reference and fields that are not
stronger than $\sim 1-3\times 10^{18}$ G in the interior we may
expect that fields of the order $10^{17}$ G could exist in the 
inner crust of the star. We consider fields in the range
$10^{16}-10^{18}$ G.

The present study is organized in the following way: in section II the
formalism are presented, in section III results are discussed and the main conclusions are drawn in section IV.
\section{Formalism} \label{form}
We describe the nuclear matter at the inner crust within  a relativistic mean-field (RMF) 
approach, in which the nucleons interact via the exchange of mesons. The
exchanged mesons are the isoscalar- scalar and vector mesons ($\sigma$ and
$\omega$, respectively) and the  isovector meson ($\rho$). We consider a system of protons and neutrons
 with mass $M$ interacting with and through an isoscalar-scalar field $\phi$ with mass $m_{s}$, an isoscalar-vector 
field $V^{\mu}$ with mass $m_{v}$, an isovector-vector field ${\bf b}^{\mu}$ with mass $m_{\rho}$. We also include a system 
of electrons with mass $m_{e}$ to obtain a charge neutral system. Protons and electrons interact through the electromagnetic 
field $A^{\mu}$.The Lagrangian density reads:

\begin{equation}\label{eq3.2}
	\mathcal{L} = \sum_{i=p,n} \mathcal{L}_{i} + \mathcal{L}_{e} + \mathcal{L}_{\sigma} + \mathcal{L}_{\omega} + \mathcal{L}_{\rho} + \mathcal{L}_{\gamma} \ \ ,  
\end{equation}
\noindent
where the nucleon Lagrangian reads

\begin{equation}
 {\cal L}_{i}  =  \bar{\psi}_{i} [\gamma_{\mu} i D^{\mu} - M^{*}] \psi_{i} ,
\end{equation}
\noindent
with
\begin{eqnarray}
	& iD^\mu = i\partial^\mu - g_{v}V^{\mu} - \frac{g_{\rho}}{2} {\vec \tau} \cdot {\bf b}^{\mu} - e \frac{1+\tau_{3}}{2} A^{\mu}  , \\
	& M^* = M - g_{s}\phi ,
\end{eqnarray}
\noindent
and the electron Lagrangian is given by
\begin{equation}
 {\cal L}_{e}  =  \bar{\psi}_{e} [\gamma_{\mu}(i\partial^{\mu} + eA^{\mu}) - m_{e} ]\psi_{e} \ , \\
\end{equation}
and the meson Lagrangian densities are
\begin{eqnarray}
 {\cal L}_{\sigma} & = &\frac{1}{2} (\partial_{\mu}\phi \partial^{\mu} \phi - m_{s}^2 \phi^2 - \frac{1}{3} \kappa \phi^3 - \frac{1}{12} \lambda \phi^4 ) \ , \\
 {\cal L}_{\omega} & = & \frac{1}{2} \left( -\frac{1}{2} \Omega_{\mu\nu} \Omega^{\mu\nu} + m_{v}^2 V_{\mu}V^{\mu} + \frac{1}{12} \xi g_{v}^4 (V_{\mu}V^{\mu})^2 \right) \ \  \\
 {\cal L}_{\rho} & = & \frac{1}{2} (-\frac{1}{2} {\bf B}_{\mu\nu} \cdot {\bf B}^{\mu\nu} + m_{\rho}^2 {\bf b}_{\mu} \cdot {\bf b}^{\mu}) \ , \\
 {\cal L}_{\gamma} & = & -\frac{1}{4} F_{\mu\nu} F^{\mu\nu} \ ,
\end{eqnarray}
\noindent
where the tensors are given by
\begin{eqnarray}\label{tensores}
	\Omega_{\mu\nu} & = & \partial_{\mu}V_{\nu} - \partial_{\nu}V_{\mu}  \ , \\
	{\bf B}_{\mu\nu} & = & \partial_{\mu}{\bf b}_{\nu}   -  \partial_{\nu}{\bf b}_{\mu}  - g_{\rho}({\bf b}_{\mu} \times {\bf b}_{\nu}) \ , \\
	F_{\mu\nu} & = & \partial_{\mu} A_{\nu} - \partial_{\nu}A_{\mu} \ \ .
\end{eqnarray}

The parameters of the model are: the nucleon mass $M$, three coupling
constants, $g_{s}$, $g_{v}$, and $g_{\rho}$, of the mesons to the
nucleons, the electrons mass $m_{e}$, the masses of the mesons
$m_{s}$, $m_{v}$, $m_{\rho}$, the electromagnetic coupling constant
$e=\sqrt{4\pi/137}$, and the self-interacting coupling constants
$\kappa$, $\lambda$, and $\xi$.

We use the sets of constants proposed for parametrizations NL3
\cite{nl3}, and TM1 \cite{tm1,Sugahara1994}. The
nuclear matter properties provided by these sets of parameters are displayed in Table
\ref{Tab:ConjDeParametros}. Both parametrizations have been fitted to
the ground-state properties of stable and unstable nuclei. TM1 includes
a quartic term involving the $\omega$-meson which allows for a softer
equation of state at larger densities. Both models have a symmetry
energy slope at saturation that is presently considered too
high. Nevertheless we have considered these two models as reference
since they have been widely used and we expect that the general features
obtained with these models will be valid for other models. 

We will not consider the effect of the anomalous magnetic moment
because its effect is only important for magnetic fields stronger than
the ones considered in the present study, \cite{broderick00,aziz08}.

\begin{table}[htdp]
\caption{Nuclear matter properties of NL3 and TM1.}\label{Tab:ConjDeParametros}
\centering
\begin{tabular}{lcc}
	\hline
	\hline
		\phantom{-------SPACE--------}			&	 NL3\cite{nl3} 		&	 TM1\cite{Sugahara1994}  \\
	\hline

	$\rho_{0}(\text{fm}^{-3})$					&		0.148			&	0.145	\\
	$M^*/M$							&		0.60				&	0.634	\\
	$\sigma(Y_{p}=0.3)\ (\text{MeV/fm}^2)$		&		0.481			&	0.492	\\	
	$\sigma(Y_{p}=0.5)\ (\text{MeV/fm}^2)$		&		1.123			&	1.077	\\	
	\hline
	\hline
\end{tabular}
\end{table}%
\section{The Thomas-Fermi Approximation}
From the Euler-Lagrange formalism, we obtain from eq.(\ref{eq3.2}) the coupled equations of motion for the scalar, isoscalar-vector,
 isovector-vector, electromagnetic, and nucleon fields. For a static system only the zero components of the vector fields and 
currents will be  present and due to charge conservation only the third component of the $\rho$-field remains.  
Therefore, the equations of motion in the RMF approximation become:
 
\begin{align}
	(-\nabla^2 + m_{s}^2)\phi({\bf r}) = & \ g_{s}\rho_{s}({\bf r}) - \frac{1}{2}\kappa \phi^2({\bf r})  - \frac{1}{6} \lambda \phi^3({\bf r})  \label{Eq.Movimento.MesonsII} \ , \\ 
	(-\nabla^2 + m_{v}^2)V_{0}({\bf r})  =  & \ g_{v} \rho_{B}({\bf r}) - \frac{1}{3!} \xi g_{v}^4 V_{0}^3({\bf r})  \label{Eq.Movimento.MesonsIII} \ , \\ 
	(-\nabla^2 + m_{\rho}^2)b_{0}({\bf r})   = & \ \frac{g_{\rho}}{2}\rho_{3}({\bf r}) \label{Eq.Movimento.MesonsIV} \ , \\
	-\nabla^2 A_{0}({\bf r})  =  & \ e(\rho_{p}({\bf r}) - \rho_{e}({\bf r})) \label{Eq.Movimento.MesonsV} \ ,
\end{align}
\noindent
where $\rho_{B}$ is the baryonic density, $\rho_{3}$ is the isospin density, $\rho_{p}$, $\rho_{n}$, and $\rho_{e}$ are 
the proton, neutron, and electron densities, and $\rho_{s}$ is the scalar density. These quantities are given by
\begin{eqnarray}\label{DensidMedias}
	\rho_{B}({\bf r}) & = &  \rho_{p}({\bf r}) + \rho_{n}({\bf r}) \ = \ <\hat{\psi}^{\dagger}\hat{\psi}> \ ,\nonumber\\
	\rho_{3}({\bf r}) & = &  \rho_{p}({\bf r}) - \rho_{n}({\bf r}) \ = \ <\hat{\psi}^{\dagger}\tau_{3}\hat{\psi}> \ , \nonumber \\
	\rho_{s} ({\bf r}) & = & \rho_{s_{p}}({\bf r}) + \rho_{s_{n}}({\bf r}) \ = \ <\hat{\bar{\psi}}\hat{\psi}> \ , \nonumber \\
	\rho_{e}({\bf r}) & = &  <\hat{\psi}_{e}^{\dagger}\hat{\psi}_{e}>   \label{DistDensidades} \ ,
\end{eqnarray}
where $\langle  \rangle$ stands for the expectation values of the field operators. 

The nucleon field operators $\hat{\psi}^{\dagger}$ and  $\hat{\psi}$ are expanded in a single particle basis which 
for infinity nuclear matter in the mean field approximation are plane wave states since the system is 
translationally invariant. In this work, as usual, the negative 
energy states will be neglected (no-sea approximation).  
  We assume that matter consists of neutrons, protons and electrons in
a strong external homogeneous magnetic field $\bf B$ in the z-direction. The gauge is fixed defining the 4-vector:
\begin{equation}
 A^{\mu}=(0,0,Bx,0)~~, 
\end{equation}
where we have $\bf B$=$B$ $\hat{z}$ and $\nabla \cdot {\bf A}$=0. 
At zero temperature all  particle densities are calculated by occupying all
 single-particle levels in the positive energy Fermi sea until the Fermi level. 
These single-particle levels are solutions of o Dirac equation where the motion is free along
the $\bf B$ field direction and quantized in the plane perpendicular to the field, yielding the 
Landau quantization\cite{landau}.
The energy dispersion relations for charged particles are modified by the presence of the strong 
magnetic field\cite{broderick00} which breaks the rotational symmetry  
and for the proton and electron are given by:  
\begin{eqnarray}
 \epsilon^p &=& \sqrt{ p_z^2 + \tilde{m}_p^2  } + g_v V_0 + \frac{1}{2} g_\rho b_0 + e A_0 ~ , \\
 \epsilon^e &=& \sqrt{ p_z^2 + \tilde{m}_e^2} - e A_0 ~ , 
\end{eqnarray}
where 
\begin{eqnarray}
 {\tilde{m}}_p^2 &=& {M^\star}^2 + 2 \nu_p eB ~ , \nonumber \\
 {\tilde{m}}_e^2 &=& {m_e}^2 + 2 \nu_e eB , 
\end{eqnarray}
\begin{equation}
  \nu_i = (n + \frac{1}{2} - \frac{1}{2} \frac{q_i}{|q_i|} \sigma_z)~~,~~i=p,e ~,~\nu_i=0,1,2...
\end{equation}
where $\sigma_z=\pm 1$ is the spin component along the magnetic field
direction, $n$=0,1,2..., $q_i$ with  i=$p$, $e$ 
stands for the electric 
charge of the proton and electron respectivelly and $\nu_i$, i=$p$, $e$ is called the Landau level. Note that 
the spin degeneracy is 1 for the $\nu$=0 Landau level and 2 for $\nu >$ 0.
Therefore, the modified density of states for a spin-1/2 charged particle becomes:
\begin{equation}
  2\int \frac{d^3 p}{(2\pi)^3} \rightarrow \sum_{\sigma_z=\pm 1} \sum_{n=0}^{\infty} 
 \int \frac{e B}{(2\pi)^2} dp_z ~=~\sum_{\nu=0}^{\infty} g_\nu
 \int \frac{e B}{(2\pi)^2} dp_z ~~,
\end{equation}
where $g_\nu=1$ for $\nu$=0 and 2 for $\nu>0$. For zero temperature and charged particles, the number and energy densities read:
\begin{eqnarray}
\rho_i &=& \frac{e B}{2\pi^2}  \sum_{\nu=0}^{\nu_{max}} g_\nu 
 p_{F,\nu}^i  ~, i=p,e \\
\epsilon_i &=&   \frac{e B}{(2\pi)^2} \sum_{\nu=0}^{\nu_{max}} g_\nu \nonumber \\
&&~~~~~~~~~ [ p_{F,\nu}^i \varepsilon_F^i +  \tilde{m}_i^2 \ln ( \frac{ p_{F,\nu}^i +
\varepsilon_F^i}{p_{F,\nu}^i} )    ]       ~,~                 
\end{eqnarray}
where, $p^i_{F,\nu}$ ,$i=p,n$, is the Fermi momentum associated with the level 
with quantum number $\nu$  and $\varepsilon_F^i$ is the corresponding 
Fermi energy (or effective chemical potential).
The Fermi momenta for the proton and electron are given by:
\begin{eqnarray}
 p^p_{F,\nu} &=& ((\varepsilon_F^p)^2 - (M^\star)^2 - 2 \nu_p e B )^{\frac{1}{2}} \\
p^e_{F,\nu} &=& ( (\varepsilon_F^e)^2 - m_e^2 - 2 \nu_e e B )^{\frac{1}{2}} ~~,
\end{eqnarray}
and the condition $p^i_{F,\nu} \geq  0 $ sets an upper limit, $\nu_{max}$, in the summations.
\begin{eqnarray}
\nu_{max} &=& \left[ \frac{ (\varepsilon_F^p)^2 - (M^\star)^2}{2 e B} \right] ~, {\rm proton} \nonumber \\
\nu_{max} &=& \left[ \frac{ (\varepsilon_F^e)^2 - m_e^2}{2 e B} \right] ~,{ \rm electron}~,
\end{eqnarray}
where $\left[x\right]$ means the largest integer smaller or equal  to $x$.
For the neutron, one obtains the standard expressions\cite{walecka},
\begin{eqnarray}
\rho_n &=& \frac{(p_F^n)^3}{3\pi^2}  \nonumber \\ 
\epsilon_n &=& \frac{1}{8\pi^2} \left[ 2 p_F^n  \varepsilon_F^{n~3} -{M^\star}^2 p_F^n \varepsilon_F^n \right.
\nonumber \\
&-& \left. {M^\star}^4 \ln (\frac{p_F^n+\varepsilon_F^n }{M^\star}) \right]  ~.
\end{eqnarray} 
%
%
In the Thomas-Fermi approximation in close analogy to the density functional formalism, we assume that the meson fields 
are sufficiently slowly-varying so that the baryons are considered to be moving in locally constant fields. 
Therefore, locally the densities 
are calculated by plane waves instead of the true position dependent single particle states. Hence,  we obtain the density 
of nucleons described by a Fermi gas with position dependent Fermi momentum.  
Energy and particle densities become  position dependent and  the Thomas-Fermi equations at T=0 are obtained from the extremization of the functional,
\begin{equation}
\Omega = E_{TF} - \sum_{i=p,n,e} \mu_i~\int d^3r ~\rho_i(\vec{r})   ~,
\end{equation}
as a function of the Fermi momenta (or equivalently function of the densities) in a complete  analogy with the density functional method.  The Lagrange multipliers $\mu_i, i=p,n,e$ are introduced in order to fix the number of particles due to species conservation. The Thomas Fermi energy is given by
\begin{equation}
	E_{TF} = \int \ \epsilon(\vec{r}) \ d^3r \,
\end{equation}
\noindent
where
\begin{align}
	\epsilon(\vec{r}) = & \sum_{i=p,n,e} \epsilon_{i}(\vec{r}) + \frac{1}{2}e(\rho_{p}-\rho_{e})A_{0}(\vec{r}) + g_{v}(\rho_{p}+\rho_{n})V_{0}(\vec{r})  \nonumber\\
	& + \frac{1}{2}g_{\rho}(\rho_{p}-\rho_{n})b_{0}(\vec{r})+ \frac{1}{2}[(\vec{\nabla}\phi)^2+m_{s}^2\phi^2]  \nonumber \\
	& +\frac{\kappa}{3!}\phi^3 + \frac{\lambda}{4!}\phi^4 - \frac{1}{2}[(\vec{\nabla}V_{0})^2+m_{v}^2 V_{0}^2]   \nonumber\\
	& - \frac{1}{4!}\xi g_{v}^4 V_{0}^4 - \frac{1}{2}[(\vec{\nabla}b_{0})^2 + m_{\rho}^2 b_{0}^2] - \frac{1}{2}[(\vec{\nabla}A_{0})^2 - B^2] \label{Eq:DensTotalEnergiafirst}\ .
\end{align}
From the condition of extremum one obtains the Thomas-Fermi equations:
\begin{eqnarray}
\mu_p &=&  \sqrt{ p_{F,\nu}^p{(\vec{r})}^{2}  +{\tilde{m}_p(\vec{r})}^2} 
\nonumber \\
 && +~g_v V_0(\vec{r})~+~\frac{1}{2} g_{\rho} b_0(\vec{r}) ~+~eA_0(\vec{r}) ~, \\
\mu_e &=& \sqrt{ p_{F,\nu}^{e}(\vec{r})^2 +\tilde{m}_e^2} ~-~eA_0(\vec{r}) ~, \\
\mu_n &=& \sqrt{p_F^{n~2}(\vec{r}) + {M^\star(\vec{r})}^2 } + g_v V_0(\vec{r}) - \frac{1}{2}g_{\rho} b_0(\vec{r})
\end{eqnarray}
In order to describe the properties of the inhomogeneous (pasta) phase we use the Wigner-Seitz approximation where the matter consisting of neutrons, protons and electrons is considered to be inside of a neutral Wigner-Seitz cell and the interaction between cells is neglected. 

Other important quantities in the study of the {\it npe} nonuniform matter are the root mean square radius $<r_{i}>\equiv\sqrt{\bar{r^2_i}}$, where
\begin{equation}
	\bar{r^2_i} = \int r^2 \rho_{i}(r)\ r^d dr /\int \rho_{i}(r) \ r^d dr \  \label{calcRaioMedio},
\end{equation}
\noindent
where $i=n,p,e$, and $d=0,1,2$ for slabs, rods or droplets respectively; the neutron skin thickness given by
\begin{equation}
	\Theta = <r_{n}> - <r_{p}> \ ,
\label{nskin}
\end{equation}
\noindent
and the surface energy defined as \cite{menezes99,pasta3}
\begin{equation}\label{Eq:SigmaSurface}
	\sigma = \int_{0}^{\infty} dr \left[ \left( \frac{d\phi_{0}}{dr} \right)^2 - \left( \frac{dV_{0}}{dr} \right)^2
 - \left( \frac{db_{0}}{dr} \right)^2\right] \ .
\end{equation}

\begin{figure}[htb]\centering
\begin{tabular}{c}
  \includegraphics[width=0.8\linewidth,angle=0]{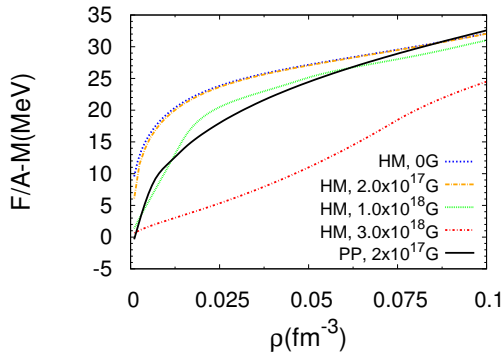}
\end{tabular}
  \caption{(Color online) Energy per particle homogeneous  $npe$ neutral matter (HM) with $Y_p=0.3$
vs density for diferent
  values of the magnetic field intensity. The calculation was
  performed with the  NL3 parametrization. For comparison the energy
  per particle obtained for the pasta phase (PP) calculation with $B=2\times
  10^{17}G$ is also included.
  }\label{free0}
\end{figure}

\begin{figure}[htb]\centering
\begin{tabular}{cc}
  \includegraphics[width=0.5\linewidth,angle=0]{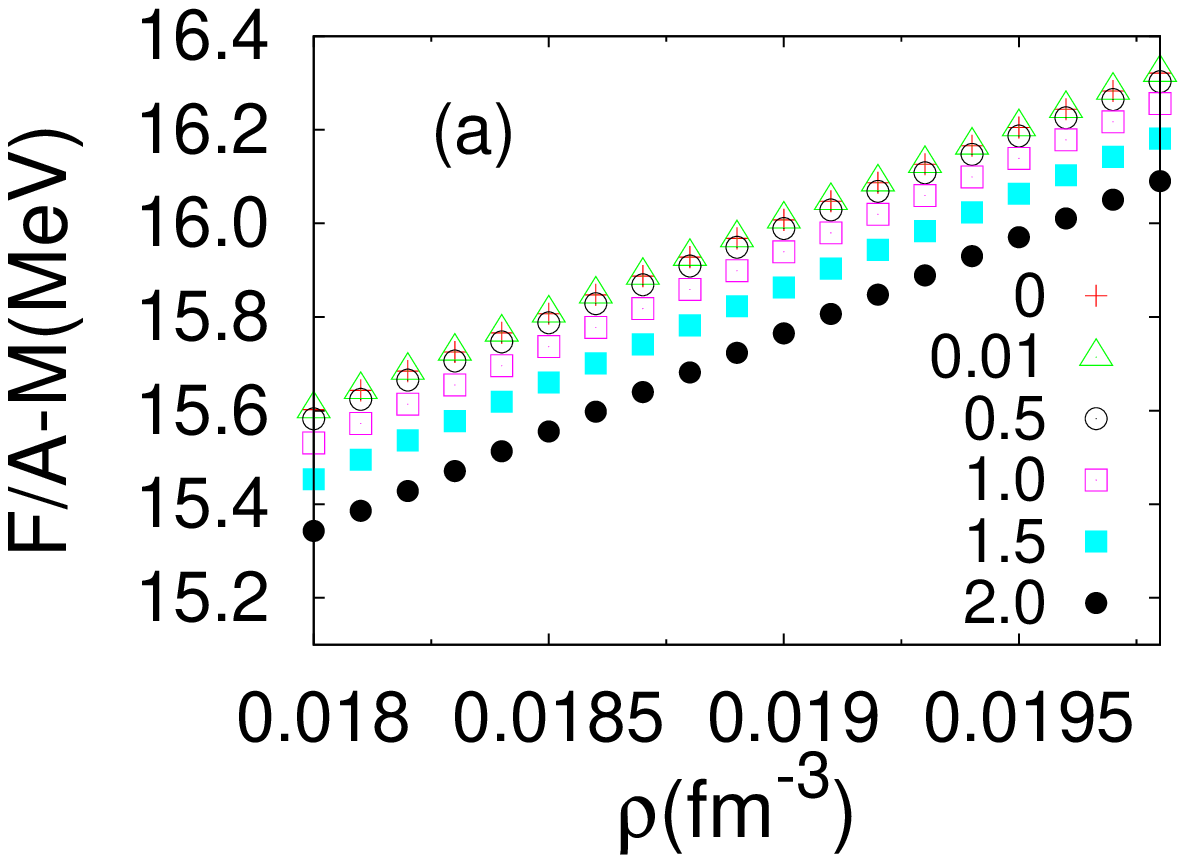}
  &
\hspace{-0.3cm}  \includegraphics[width=0.5\linewidth,angle=0]{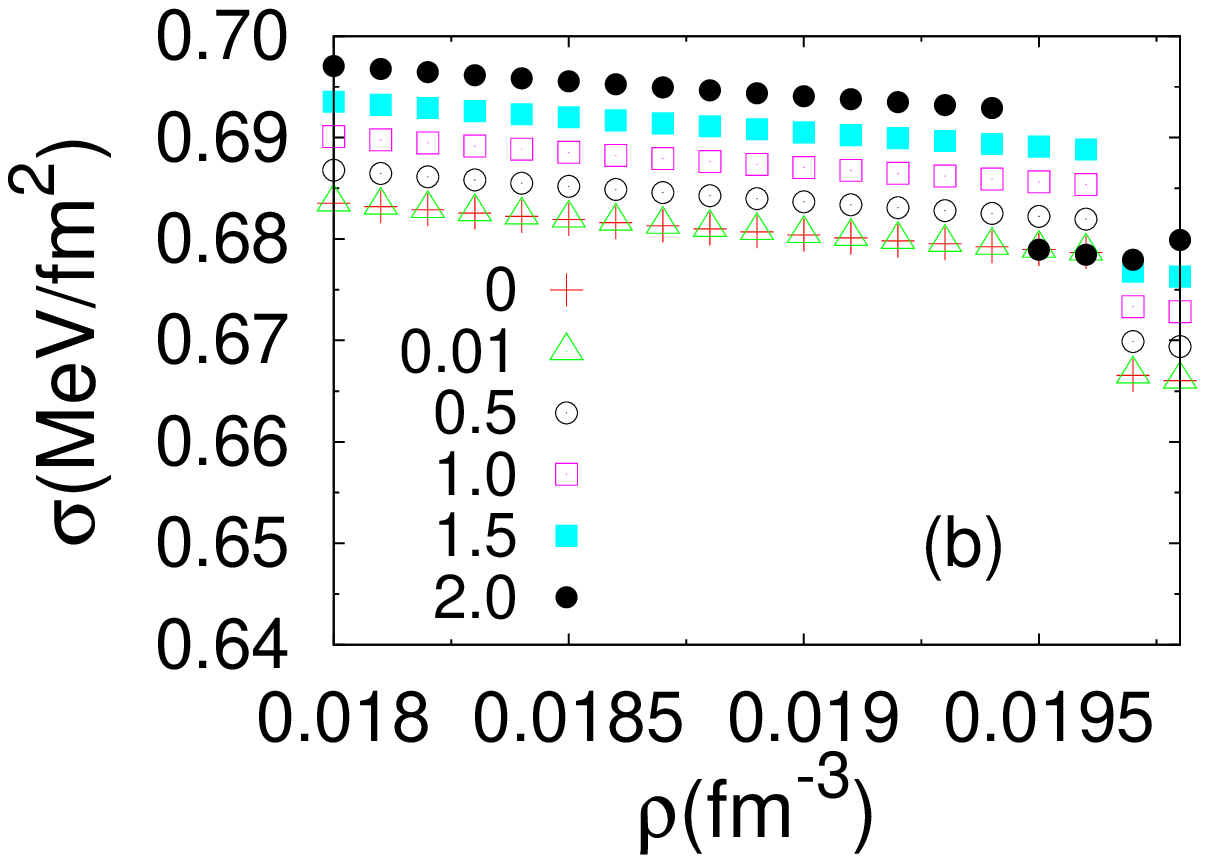} \\
 \includegraphics[width=0.5\linewidth,angle=0]{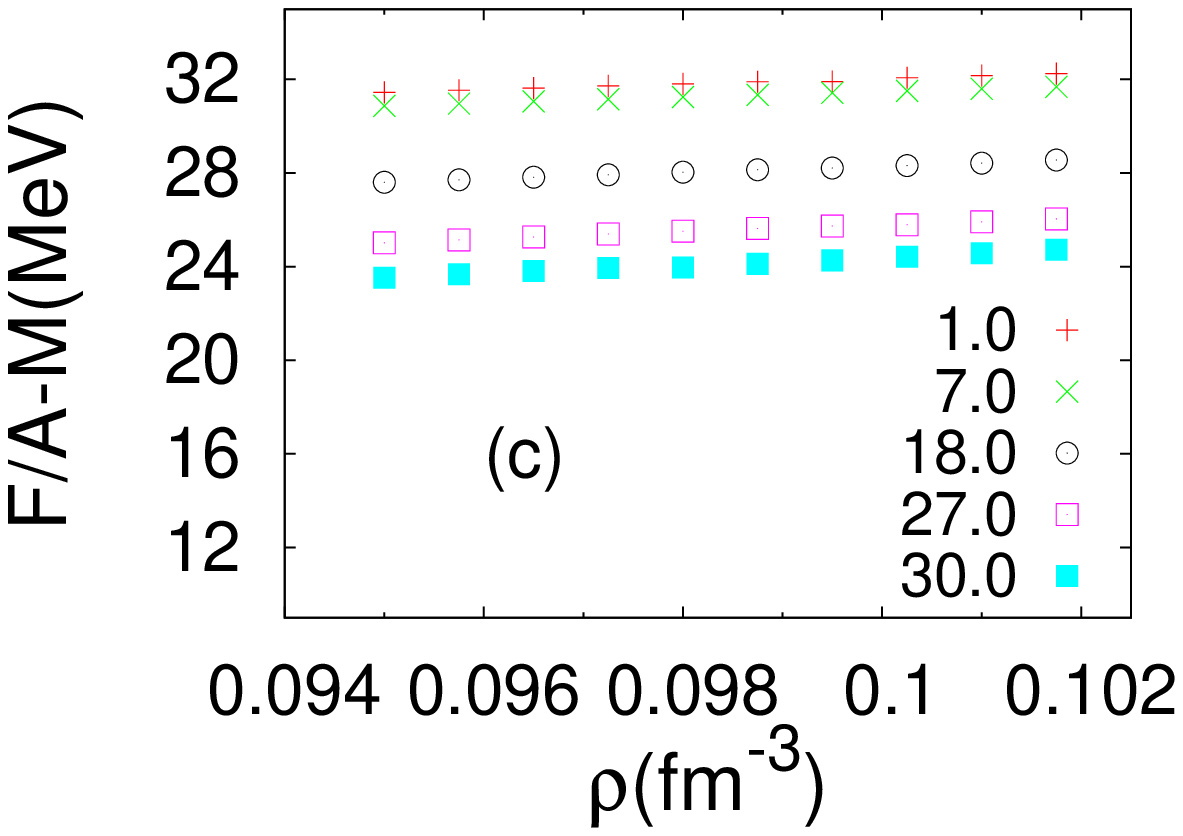} &
 \hspace{-0.3cm} \includegraphics[width=0.5\linewidth,angle=0]{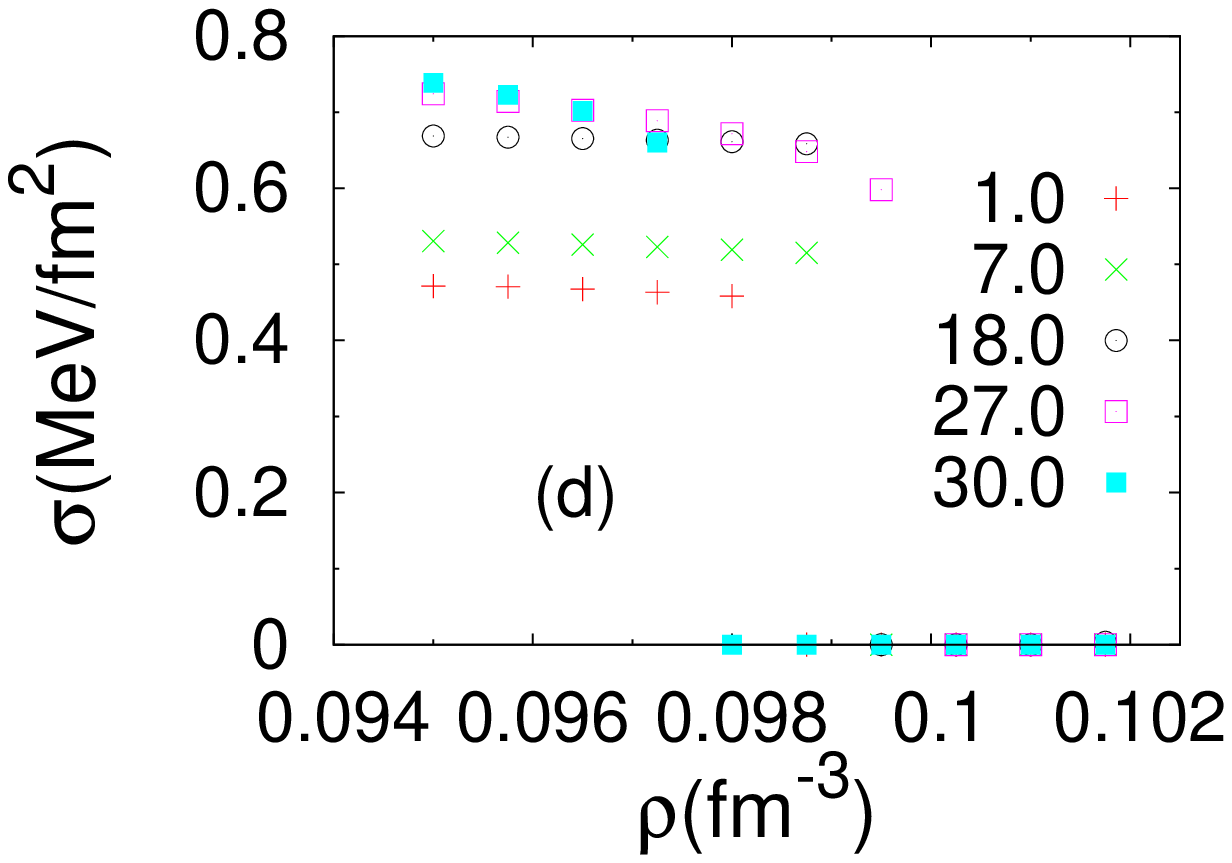} 
\end{tabular}
  \caption{(Color online) Free energy per particle (left) and surface energy (right) for
  the  NL3 parametrization and $Y_p=0.3$ at the droplet-rod transition
  (a,b)  and the bubble-homogeneous transition (c,d).  The calculations were performed for diferent
  values of the magnetic field intensity ($B$). The $B$ units are $10^{17}G$. Here the {\it free energy} is the energy itself as the system is assumed at temperature $T=0$.}\label{free}
\end{figure}

\section{Results and discussions}\label{results}

In the following we discuss the effect of the magnetic field on
several properties of the pasta clusters.  We will consider
electrically neutral matter with a fixed fraction of protons. For most
of the examples we  consider the fraction $Y_p=0.3$ a
reference value in supernova matter or proton-neutron matter, but we will also show
results for 
$Y_p=0.1$ a typical value of $\beta$-equilibrium neutron star matter. The
magnetic field is along  the rods' axis in the rod geometry, and for the slab
geometry in a direction perpendicular to the slab thickness.

\begin{figure}[htb]\centering
\begin{tabular}{ccc}
\includegraphics[width=0.5\linewidth,angle=0]{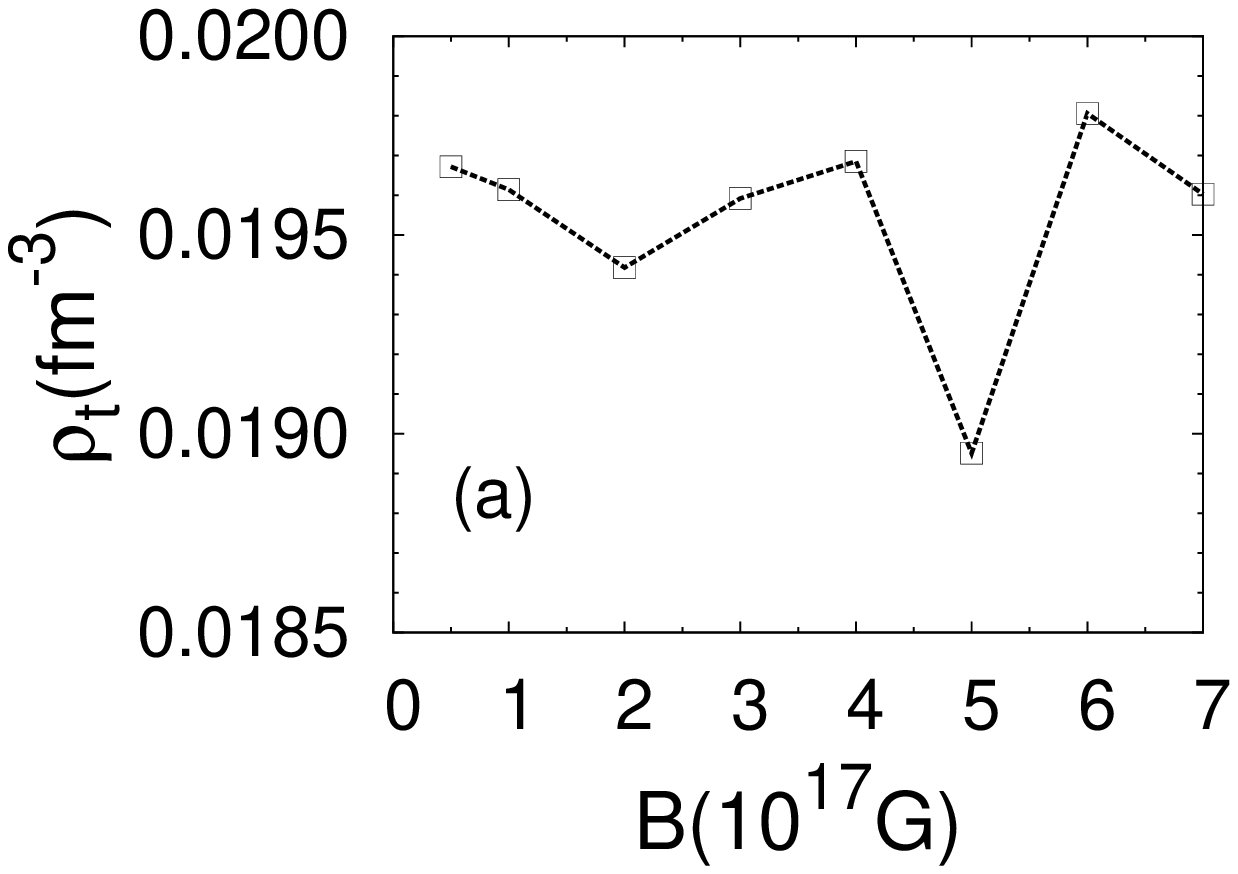} &
\includegraphics[width=0.5\linewidth,angle=0]{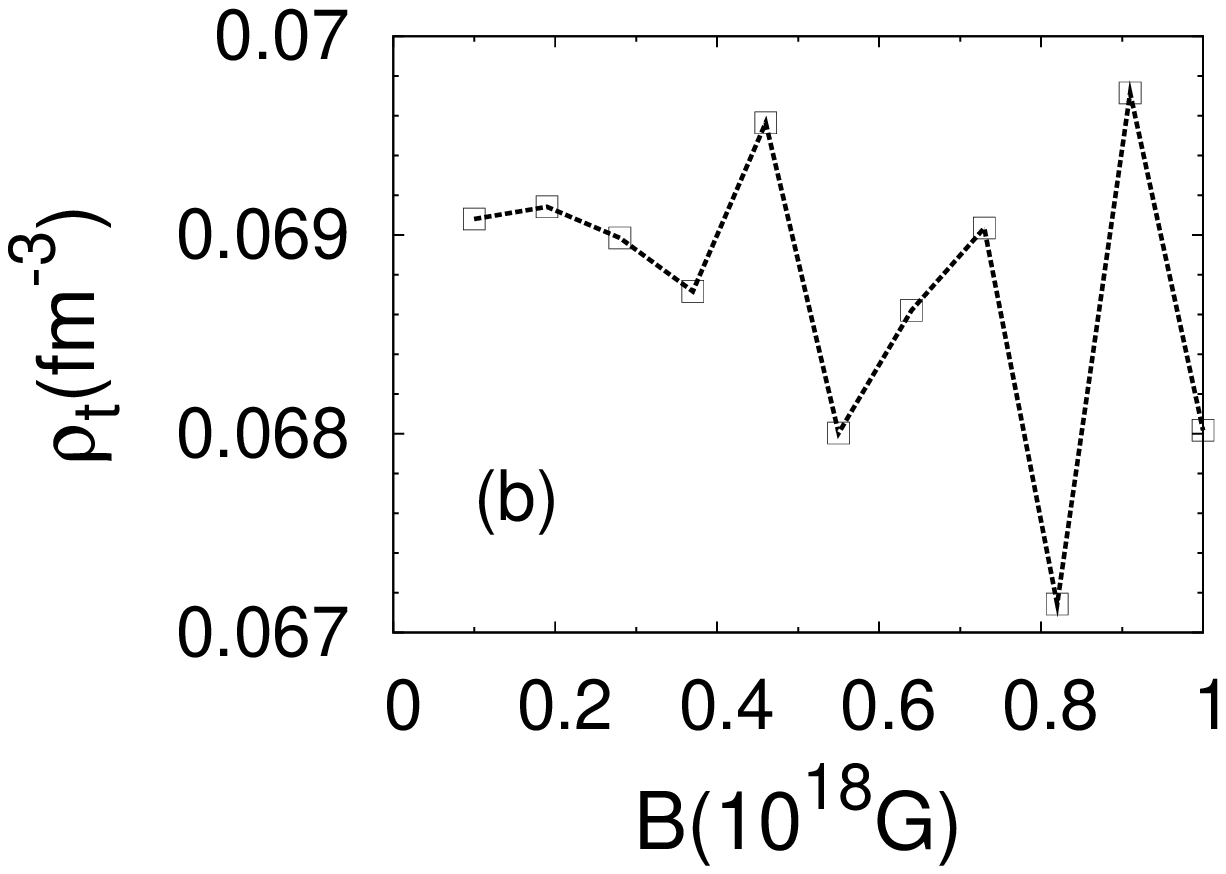} &
\end{tabular}
	\caption{Transition density for NL3  and $Y_{p}=0.3$: (a) droplet-rod; (b) 
          tube-bubble.}\label{rhot}
\end{figure}

\begin{figure}[htb]
\centering
\begin{tabular}{c}
\includegraphics[width=0.8\linewidth,angle=0]{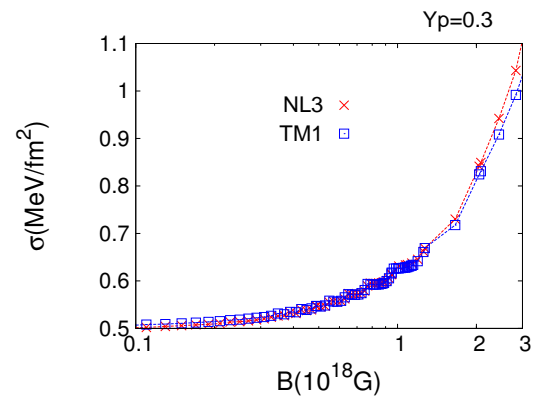} \\
\end{tabular}
\caption{(Color online) Surface energy defined by the equation (\ref{Eq:SigmaSurface}) as a function of the magnetic field for  $Y_p=0.3$, obtained with NL3 and  TM1 parametrizations and the slab configuration with the Coulomb interaction switched off for $\rho=0.06$ fm$^{-3}$.}
\label{sigma}
\end{figure}

\begin{figure}[htb]\centering
\begin{tabular}{ccc}
\includegraphics[width=0.5\linewidth,angle=0]  {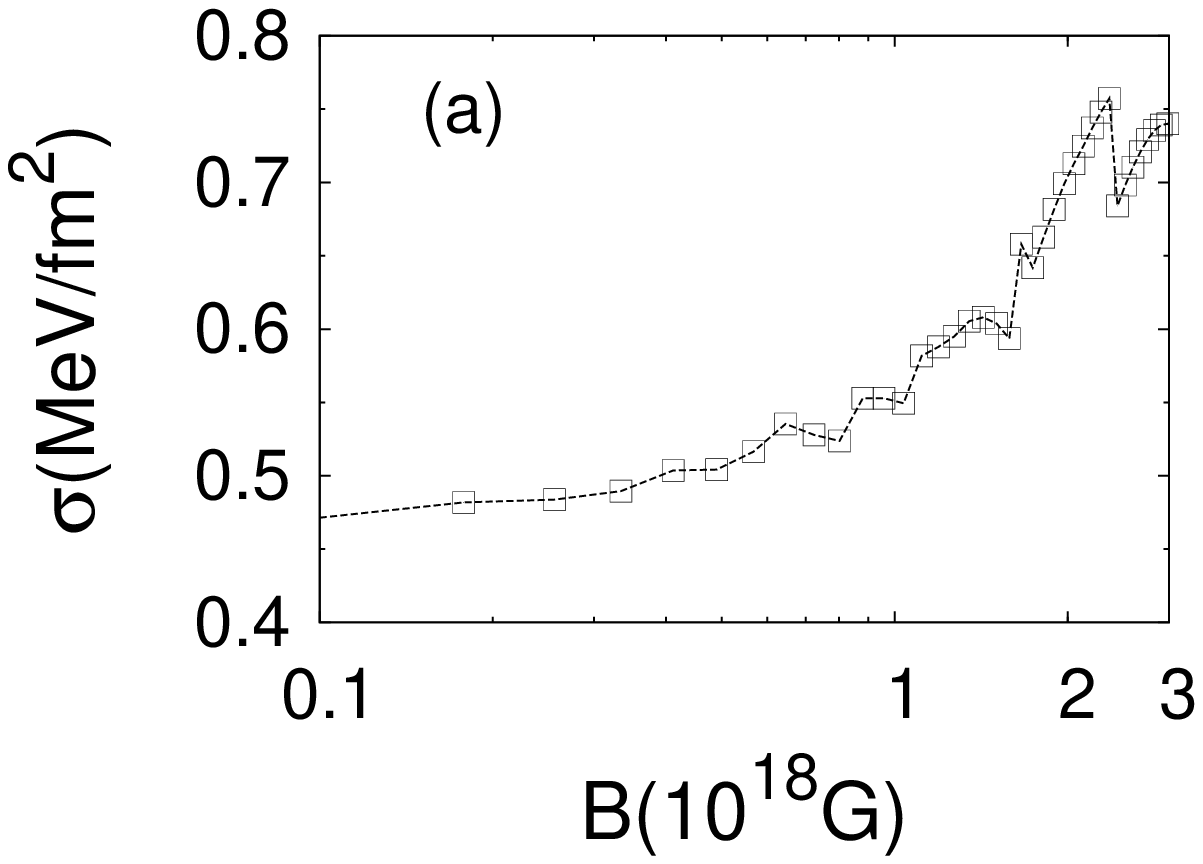} &
\includegraphics[width=0.5\linewidth,angle=0]  {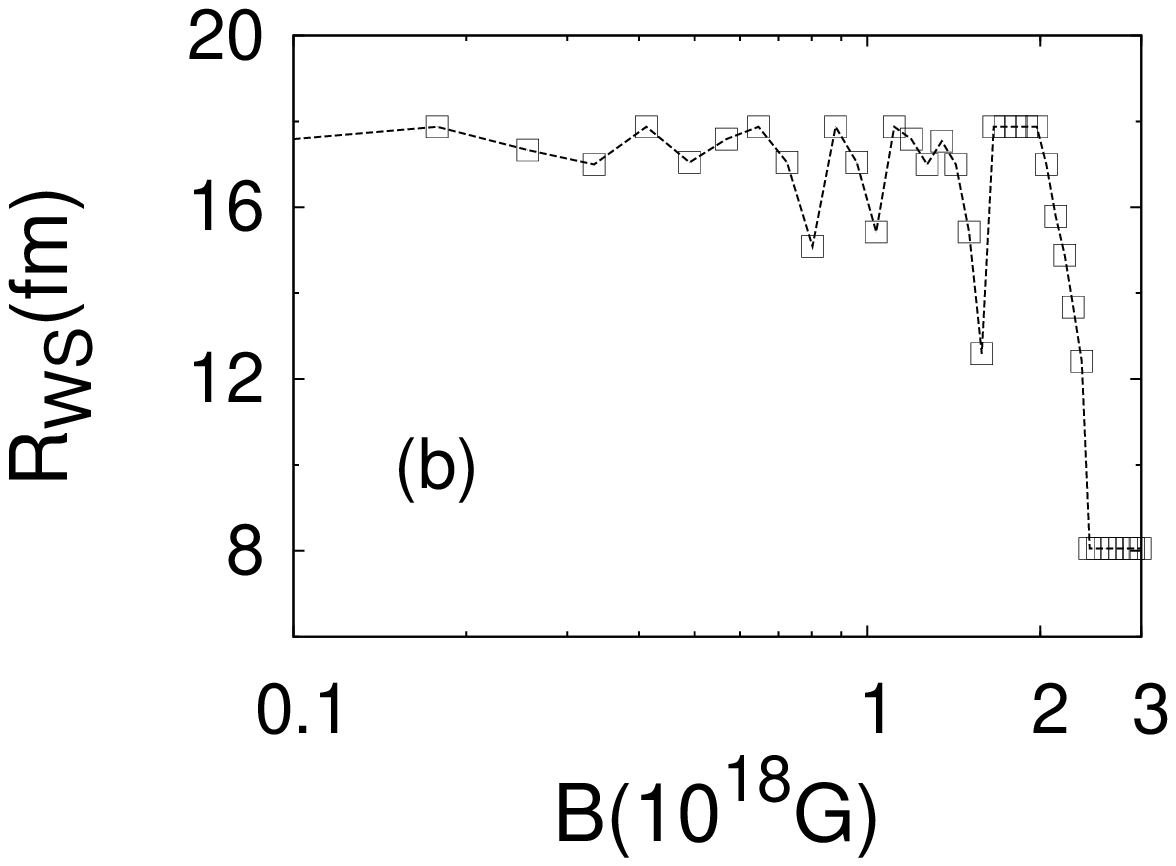} \\
\end{tabular}
	\caption{Calculations for the bubble geometry,  NL3 parametrization, $Y_{p}=0.3$ and $\rho_{B}=0.095$ fm$^{-3}$: (a) Surface energy; (b) Wigner-Seitz radius.}\label{ARwzSigma}
\end{figure}

\begin{figure}[htb]
\centering
\begin{tabular}{cc}
\includegraphics[width=0.48\linewidth, height=0.35\linewidth,angle=0]{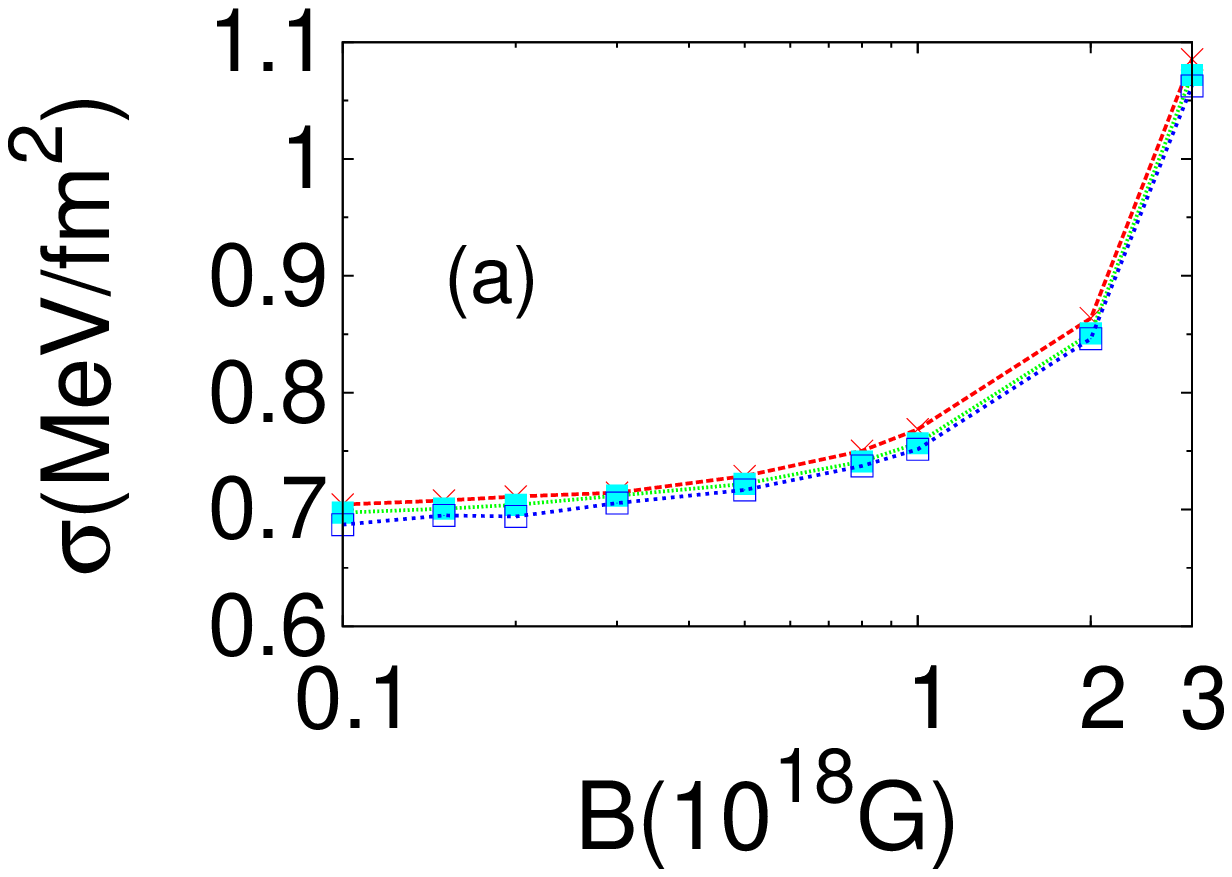} &
\includegraphics[width=0.5\linewidth,angle=0]{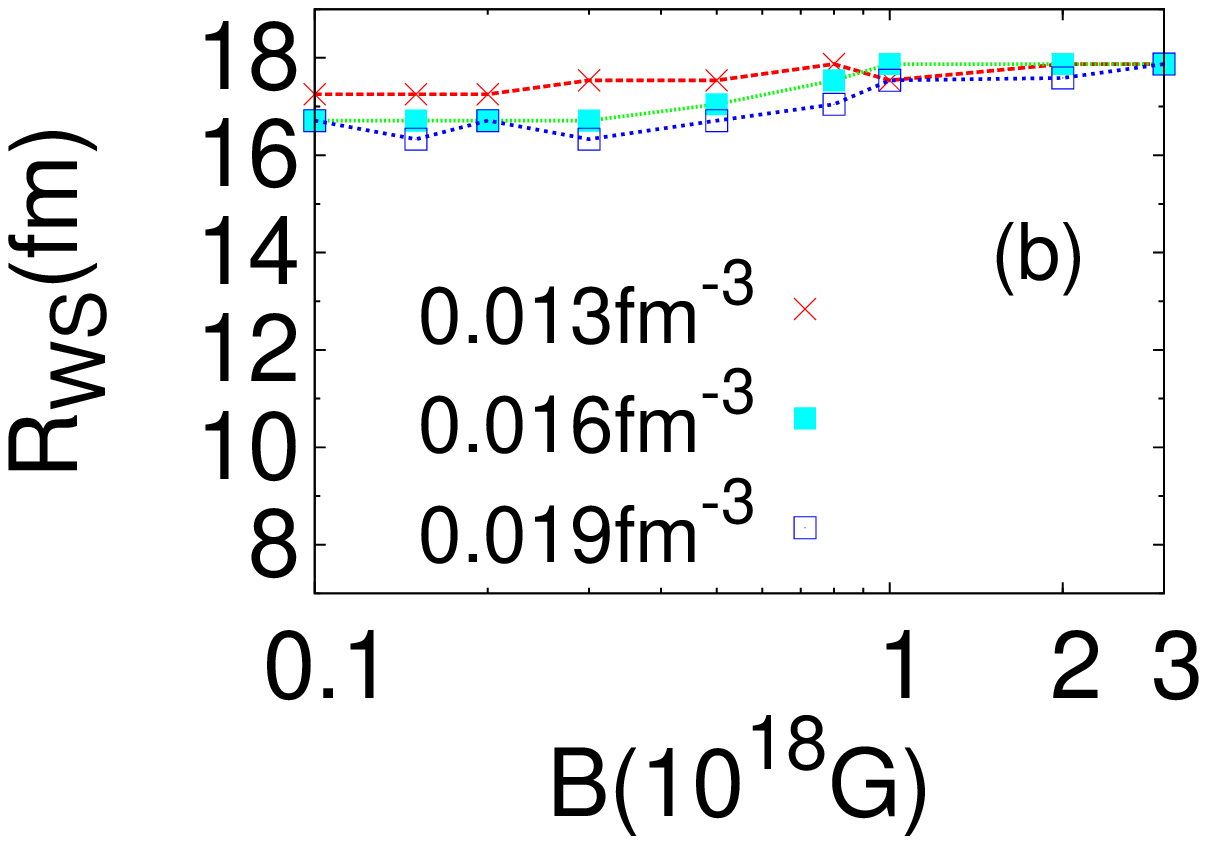} \\
\includegraphics[width=0.48\linewidth,height=0.35\linewidth,angle=0]{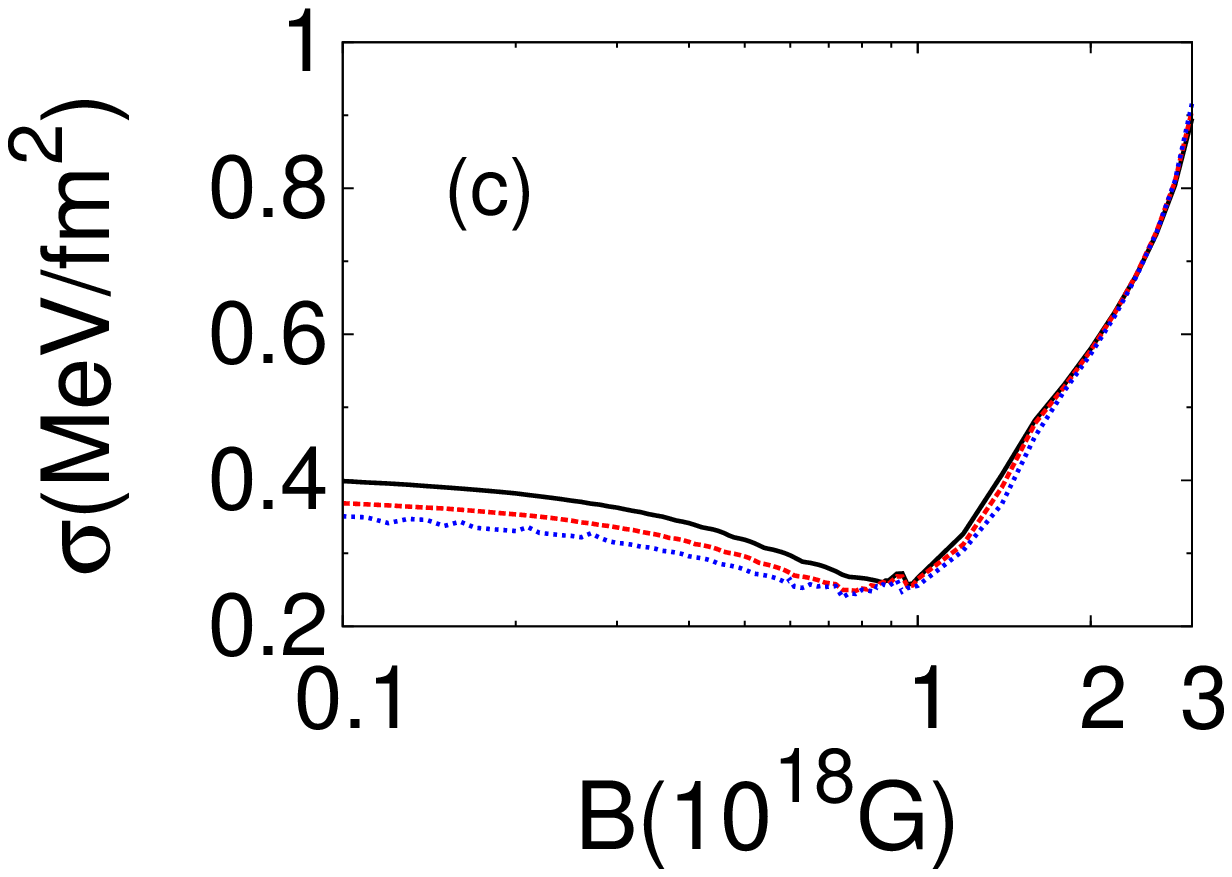} &
\includegraphics[width=0.5\linewidth,angle=0]{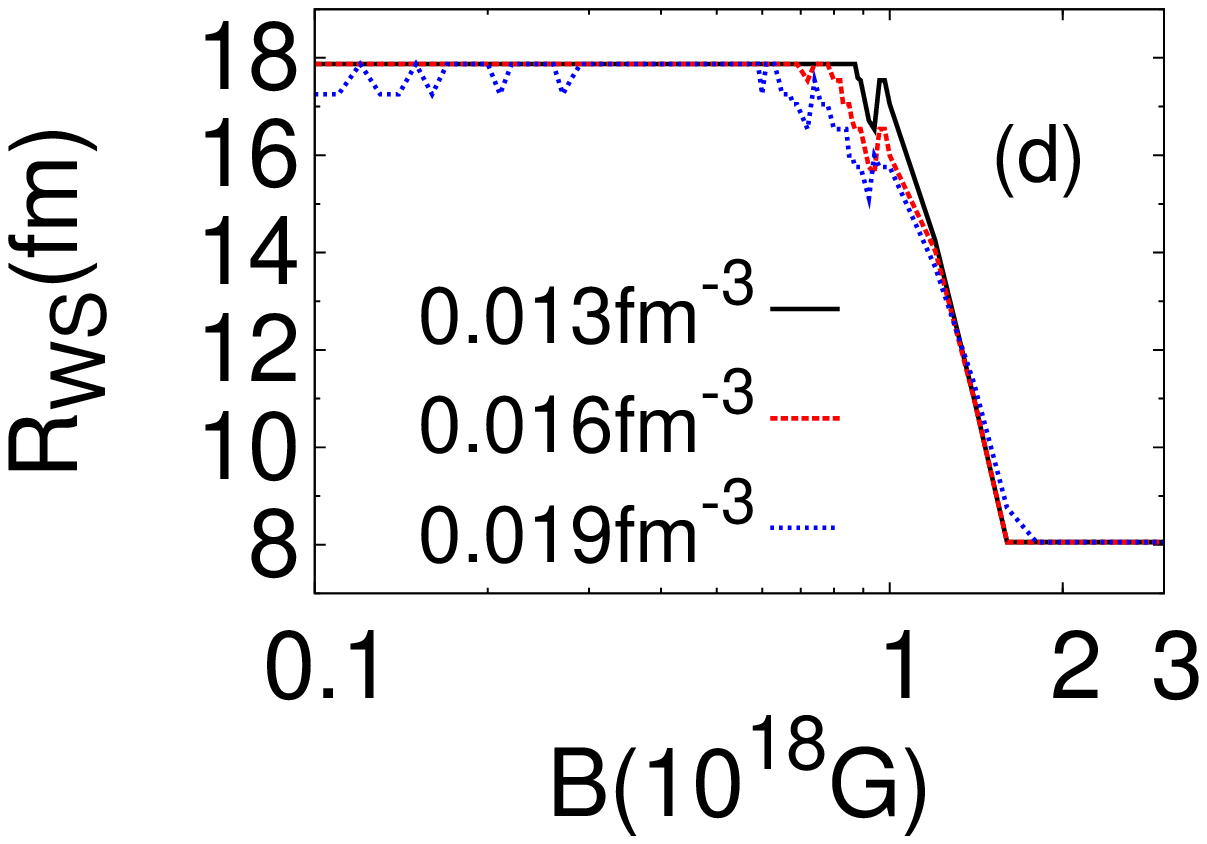}\\
\end{tabular}
\caption{(Color online) Surface energy (left) and Wigner Seitz radius (right), for the
          droplet geometry, NL3 parametrization, baryonic density $\rho=0.019$
          fm$^{-3}$ and proton fraction: (a,b) $Y_{p}=0.3$; (c,d) $Y_{p}=0.1$.}
\label{sigma2}
\end{figure}

\begin{figure}[htb]
\centering
\begin{tabular}{cc}
\includegraphics[width=0.48\linewidth,height=0.35\linewidth,angle=0] {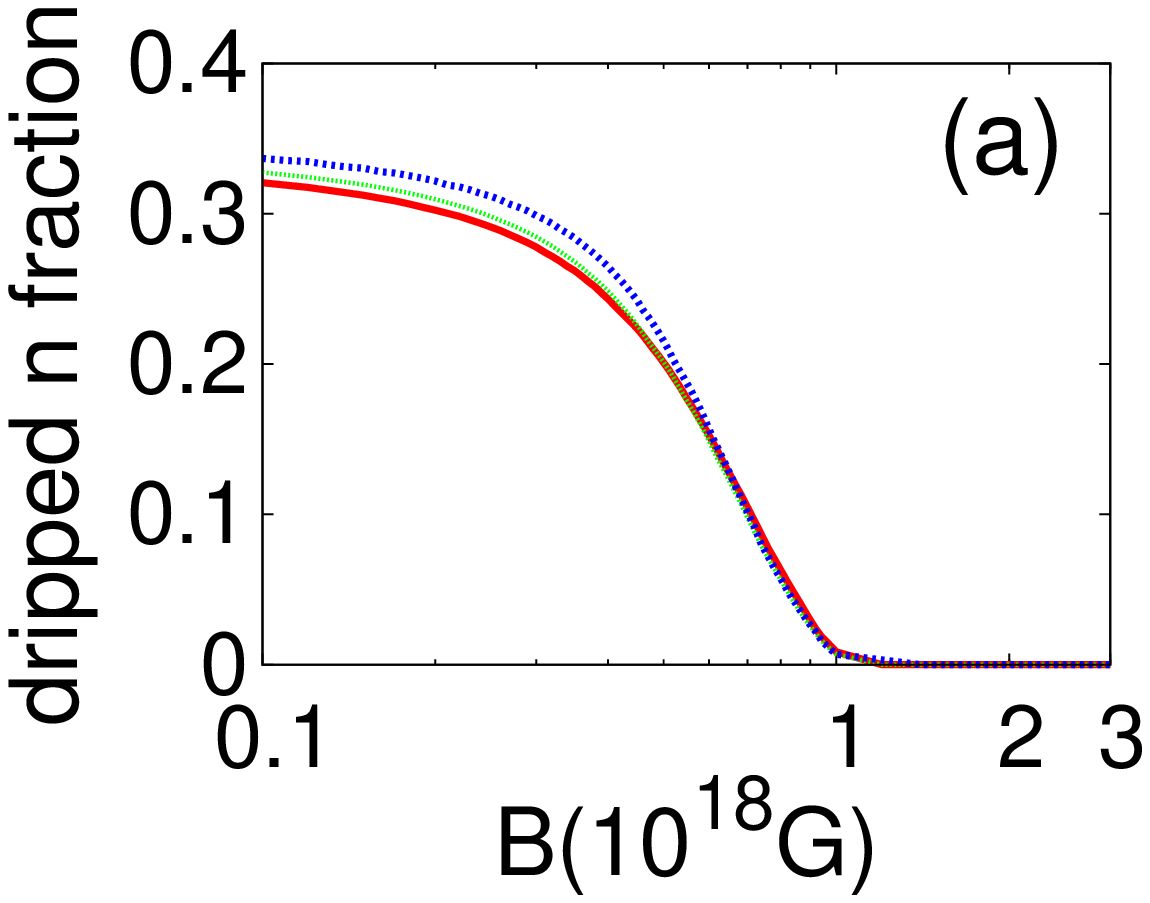} &
\includegraphics[width=0.5\linewidth,angle=0] {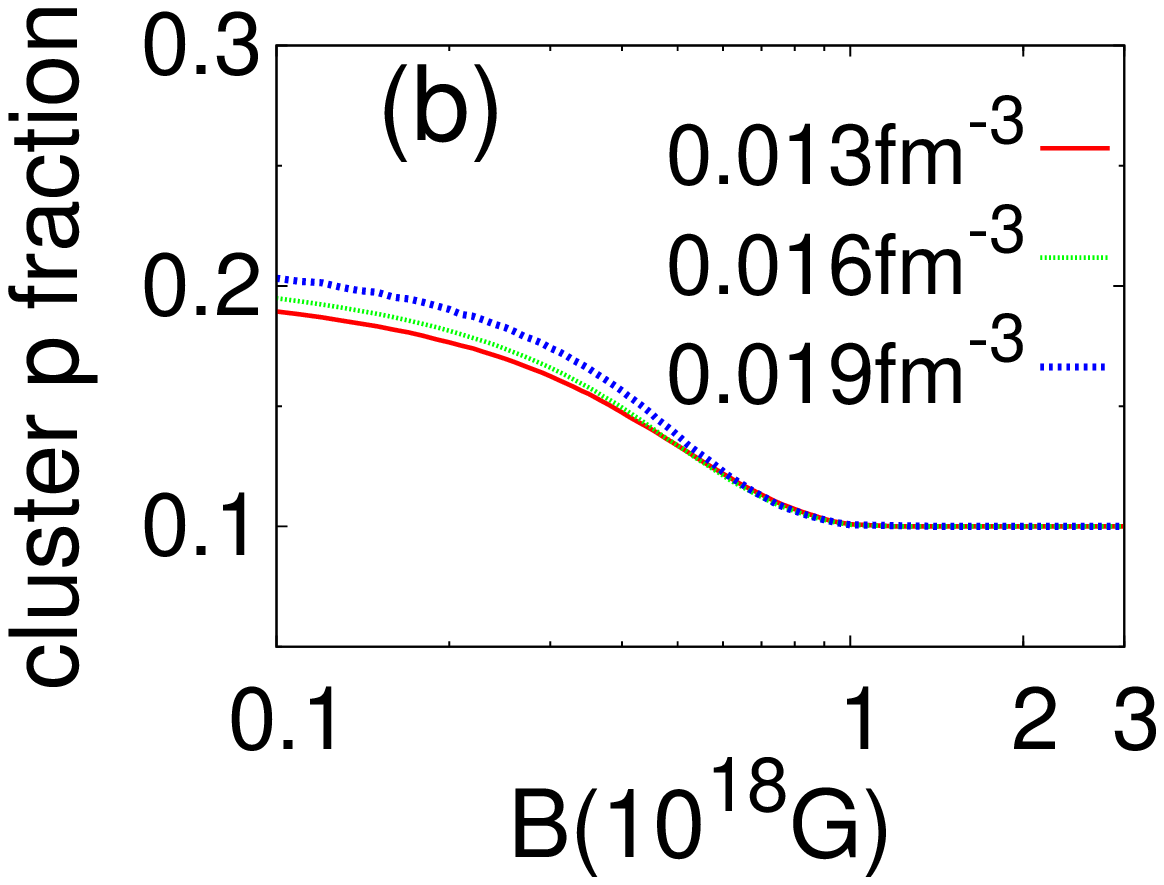}  \\
\end{tabular}
\caption{(Color online) Results in the pasta cluster for $Y_p=0.1$: (a) Dripped neutron
  fraction; (b) proton fraction.}
\label{yp01}
\end{figure}

It has been shown previously that Landau quantization softens the
equation of state (EOS) due to the large degeneracy of the Landau
levels \cite{broderick00,aziz09}.  Therefore, we expect that the free
energy per particle will decrease in the presence of an external
strong magnetic field.
 This is illustrated in  Fig. \ref{free0}  and Fig. \ref{free}, where the free
energy per particle is given as a function of the density for different
values of $B$. The homegenous matter free energy per particle is
plotted as function of the density for different values of the
magnetic field, (Fig. \ref{free0}). For reference. we include in this
same figure the results obtained within a pasta phase calcualtion for
$B=2\times 10^{17}$ G, indicating that the free energy is lower and
therefore, that this configuration is favored. We conclude that the
free energy per particle decreases when the magnetic field intensity
increases and that in this range of densities  non-homogeneous matter
is favoured.
For the
pasta calculation (Fig.\ref{free}), two density ranges have been chosen: densities close to
the drop-rod transition and the bubble-core transition. For $B=5\times
10^{16}$ G, the effect, of the order
$0.01\%$, is negligible. However, for $B=2\times 10^{17}$ G the free
energy is 2\% lower than for magnetic field free configurations. At the configuration
transition the free energy is continuous, however, the surface energy defined
by eq. (\ref{Eq:SigmaSurface}) suffers a jump.
At the crust-core transition it goes to zero while at
the drop-rod transition it suffers a small decrease. This discontinuity
is possibly  not
due to the presence of a magnetic field but to the limitation
of the calculation that only considers configurations with well defined
symmetries while intermediate geometries and shapes are expected to exist \cite{pais2012,dorso2012}. However, the
magnetic field may change  the transition density. This does not show
a systematic trend, reflecting the filling of Landau levels
and suffering a larger effect for larger magnetic fields.

\begin{figure}[htb]\centering
\begin{tabular}{ccc}
\includegraphics[width=0.5\linewidth,angle=0]{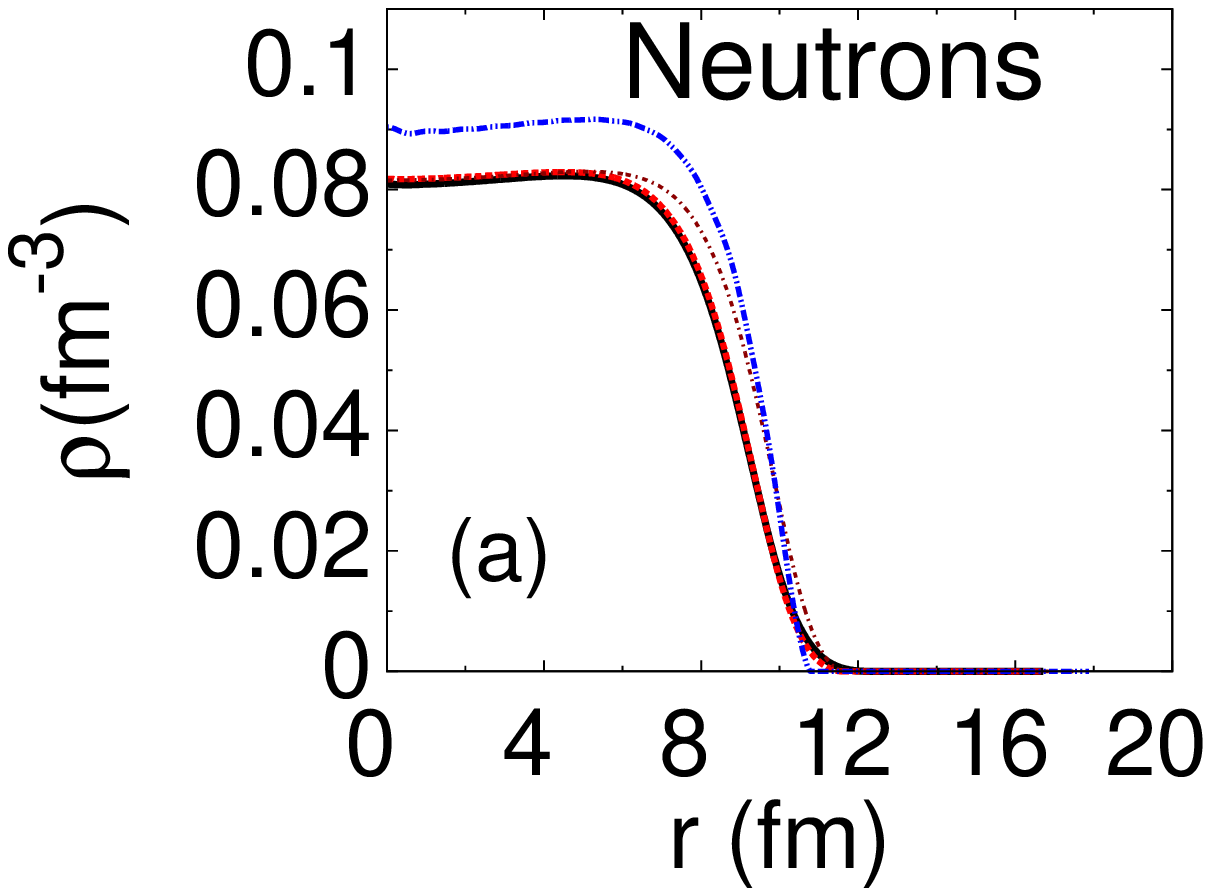} &
\includegraphics[width=0.5\linewidth,angle=0]{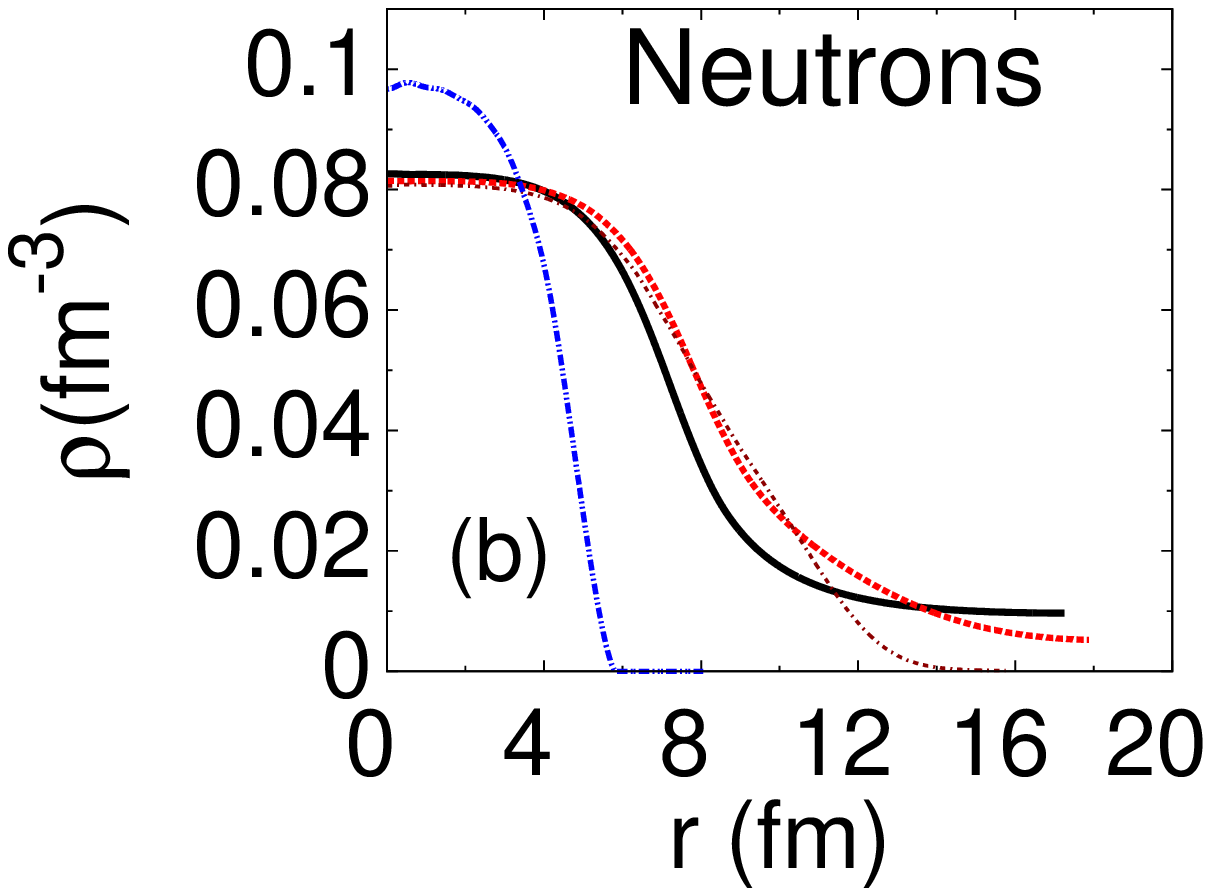} \\
\includegraphics[width=0.5\linewidth,angle=0]{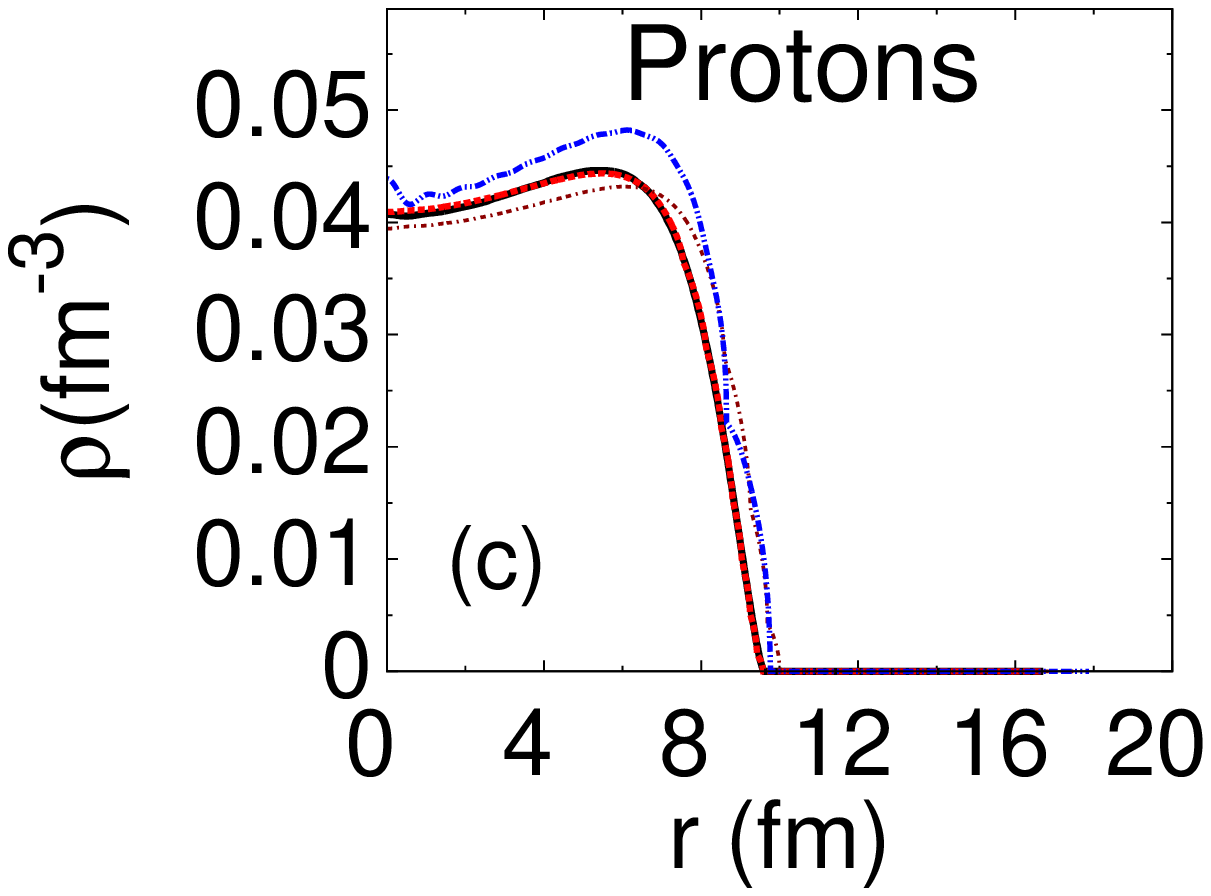} &
\includegraphics[width=0.5\linewidth,angle=0]{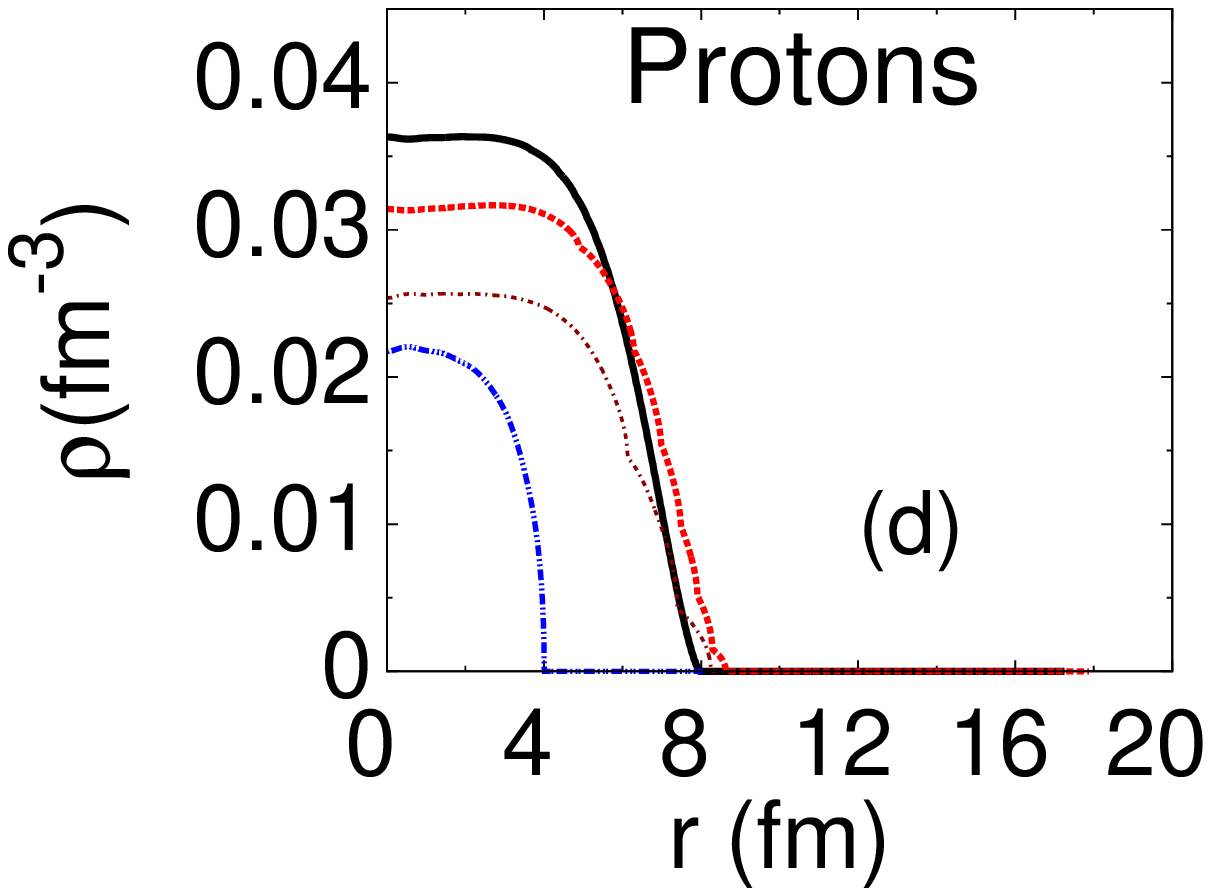} \\
\includegraphics[width=0.5\linewidth,angle=0]{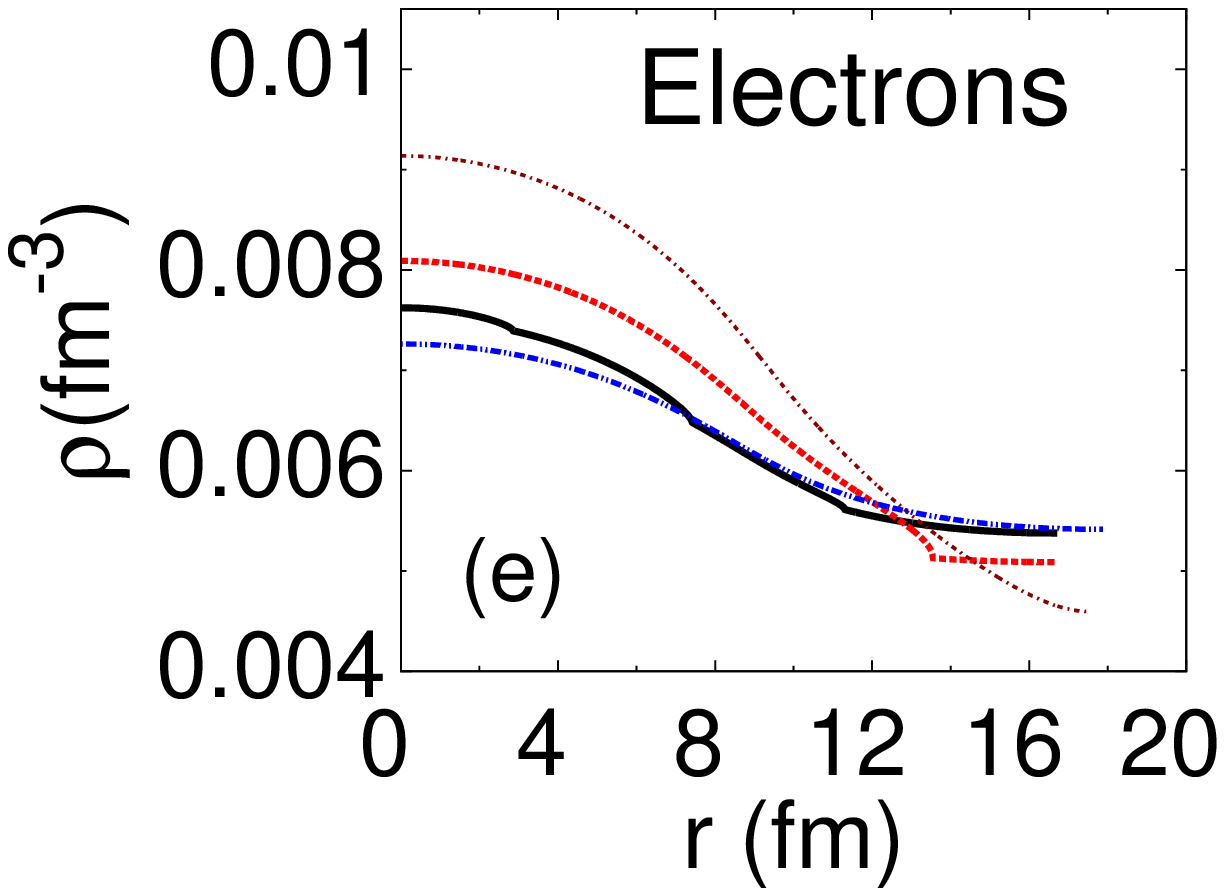} &
\includegraphics[width=0.5\linewidth,angle=0]{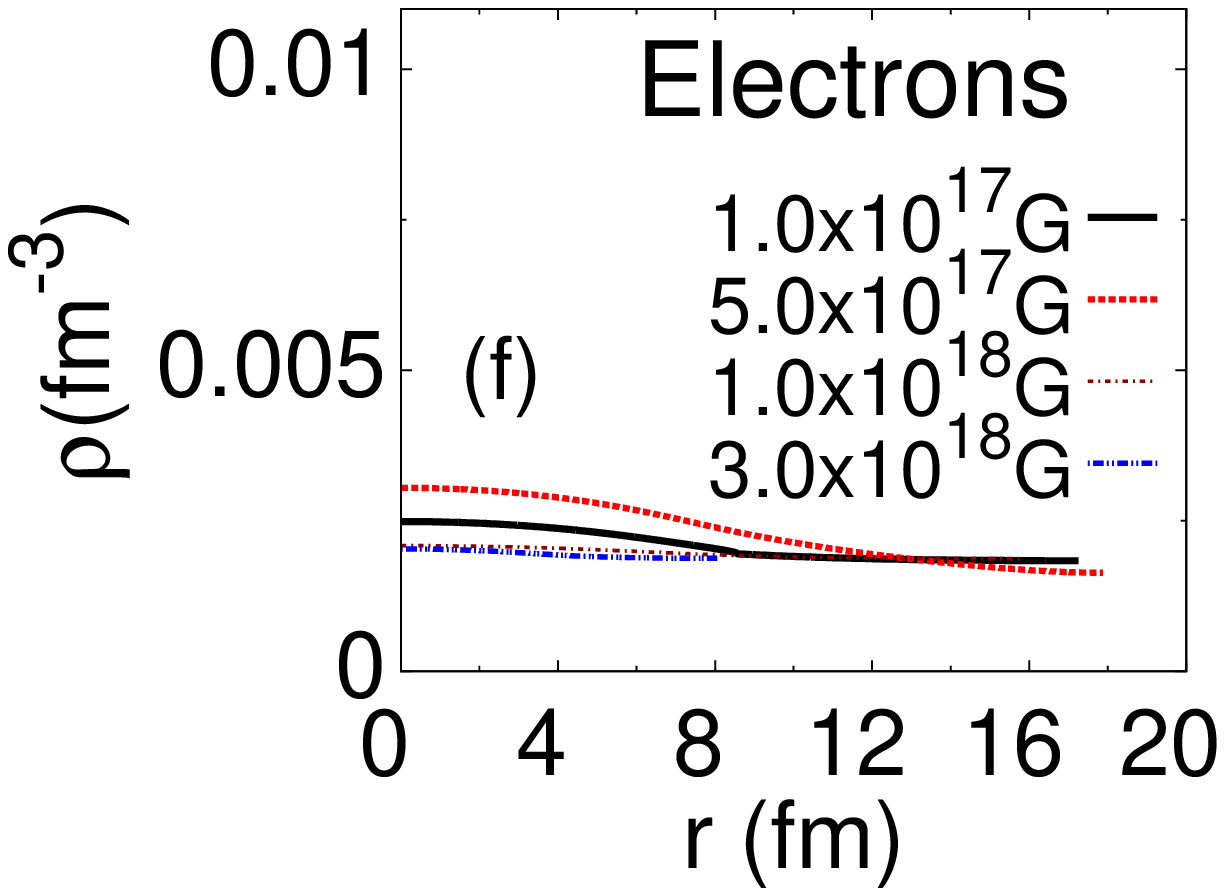} \\
\end{tabular}
	\caption{(Color online) Density profile in the Wigner-Seitz cell for the
          droplet geometry, NL3 parametrization, proton fraction
          $Y_{p}=0.3$ (left) and $Y_p=0.1$ (right) for baryonic density $\rho=0.019$ fm$^{-3}$: 
          (a,b) neutrons; (c,d)
        protons; (e,f) electrons.}\label{DensidadeDrop0.3}
\end{figure}

\begin{figure}[htb]\centering
\begin{tabular}{cc}
\includegraphics[width=0.5\linewidth,angle=0]{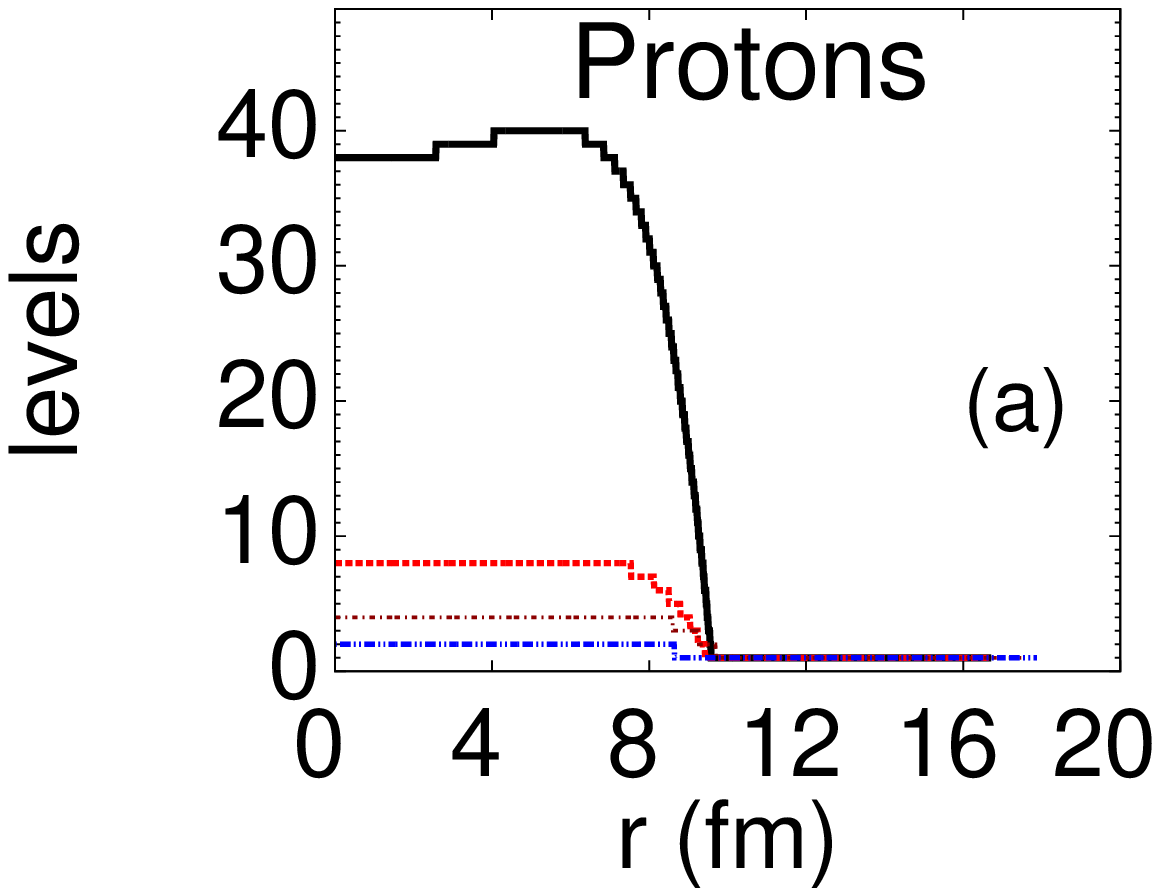} &
\includegraphics[width=0.5\linewidth,angle=0]{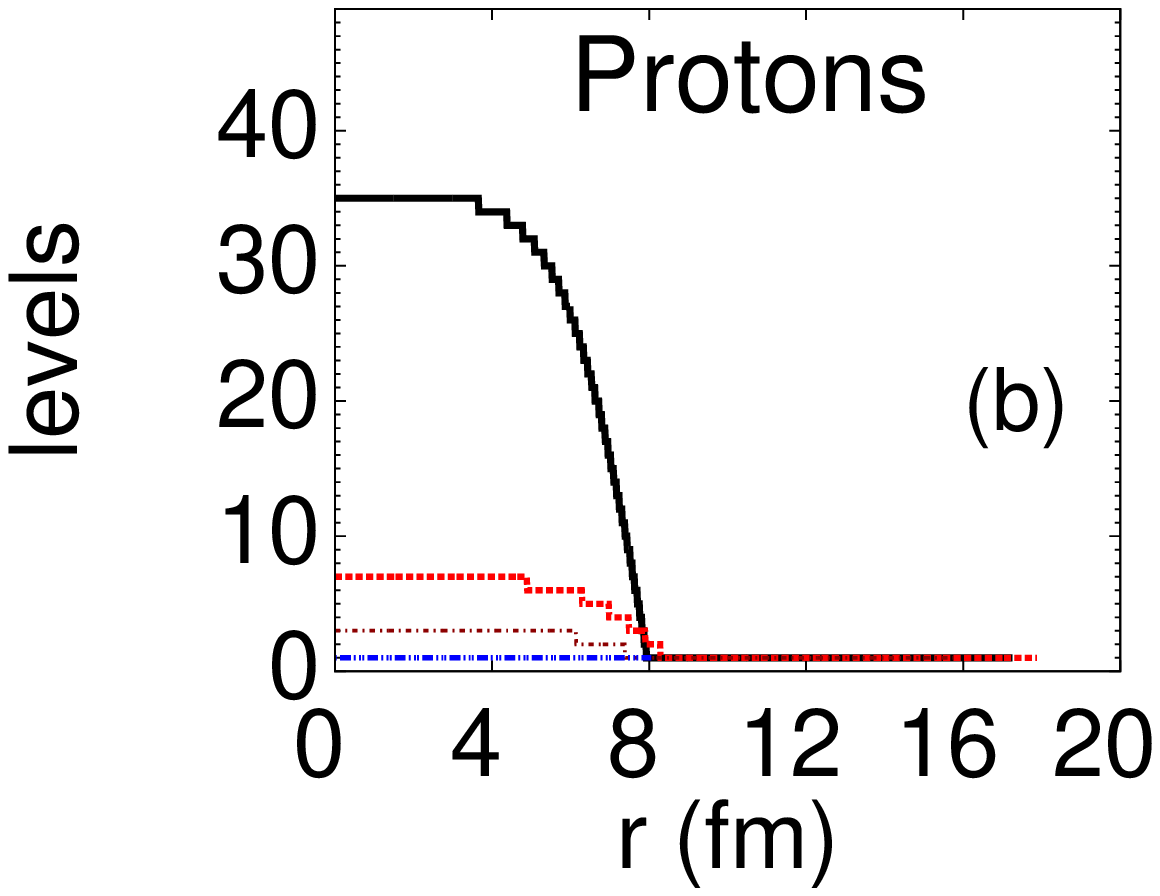} \\
\includegraphics[width=0.5\linewidth,angle=0]{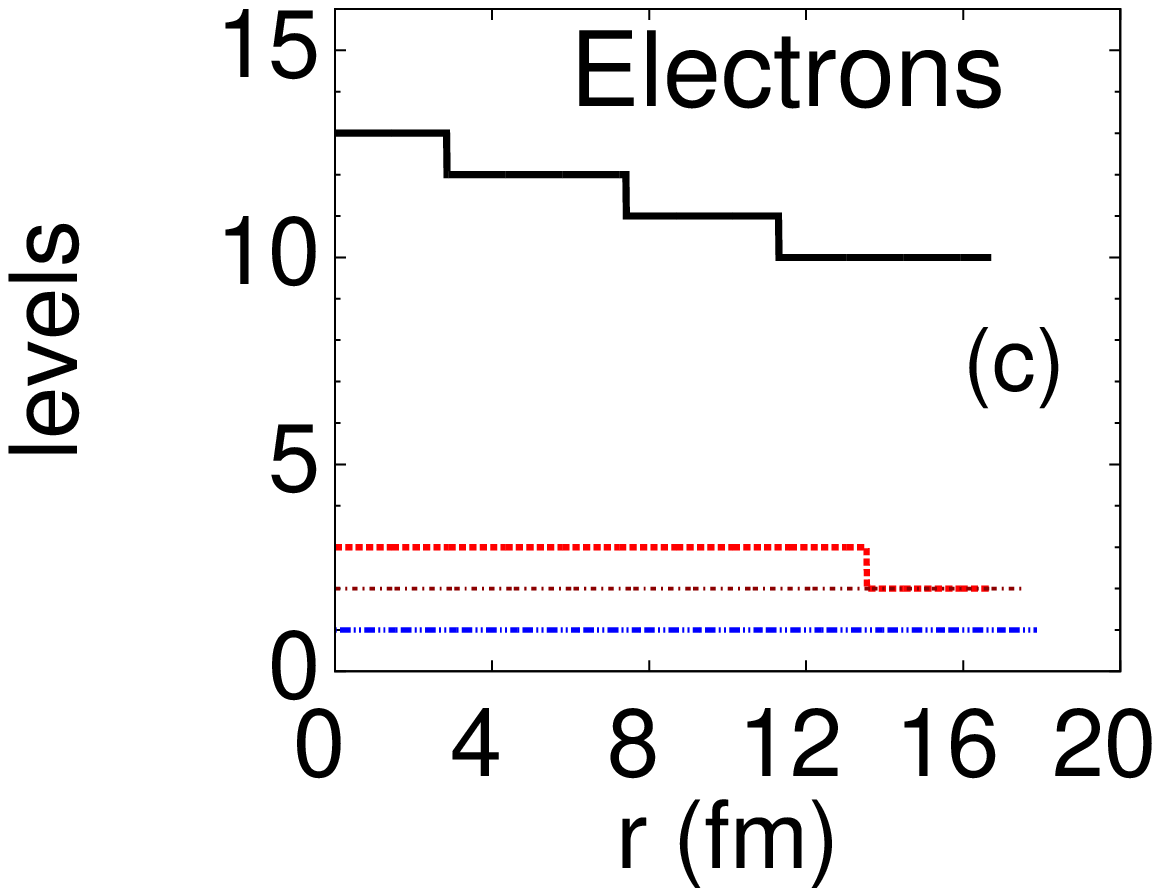}&
\includegraphics[width=0.5\linewidth,angle=0]{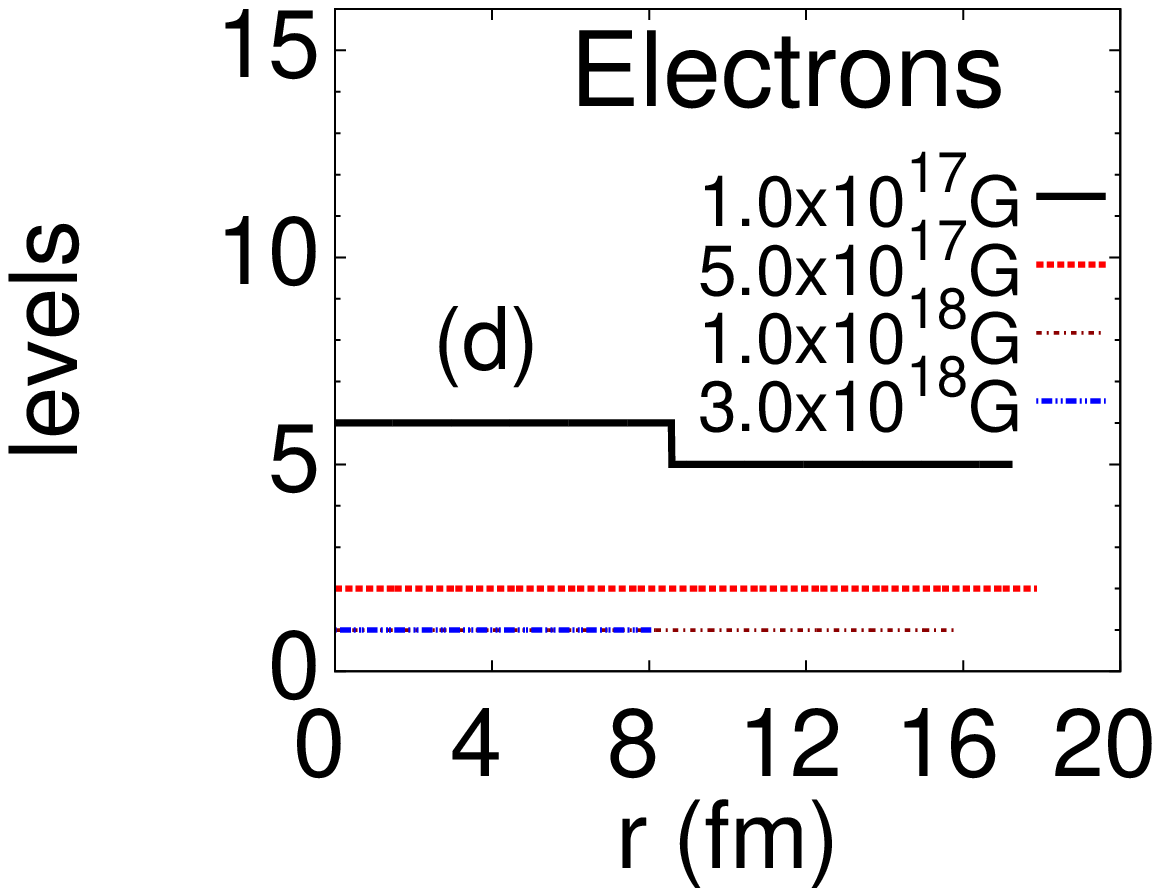} \\
\end{tabular}
	\caption{(Color online) Landau levels in the Wigner-Seitz cell for the
          droplet geometry, NL3 parametrization, proton fraction
          $Y_{p}=0.3$ (left) and  $Y_{p}=0.1$ (right) for the baryonic density
          $\rho=0.019$ fm$^{-3}$: (a,b)  protons; (c,d) electrons.}\label{LandauDrop0.1}
\end{figure}

\begin{figure}[htb]\centering
\begin{tabular}{cccc}
\includegraphics[width=0.48\linewidth,height=0.35\linewidth,angle=0]  {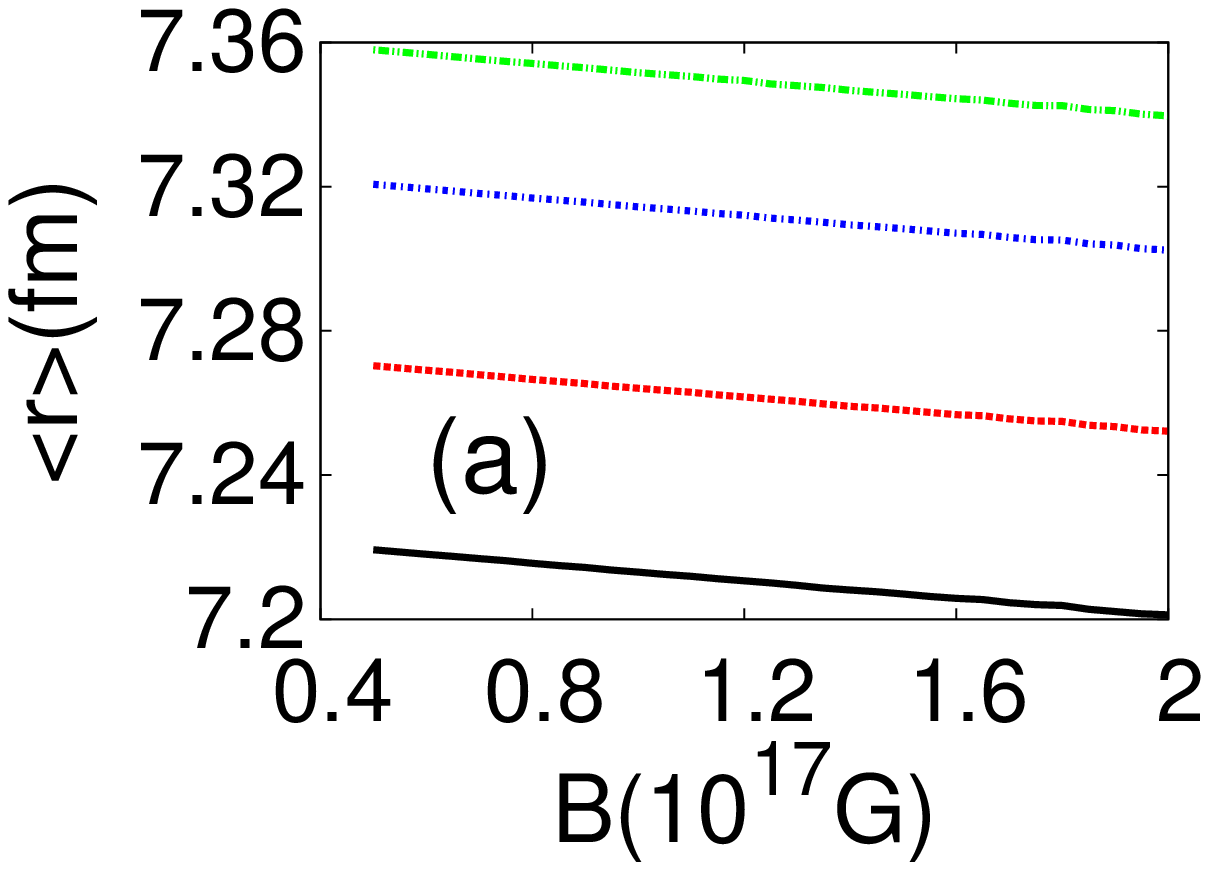} &
\includegraphics[width=0.5\linewidth,angle=0]  {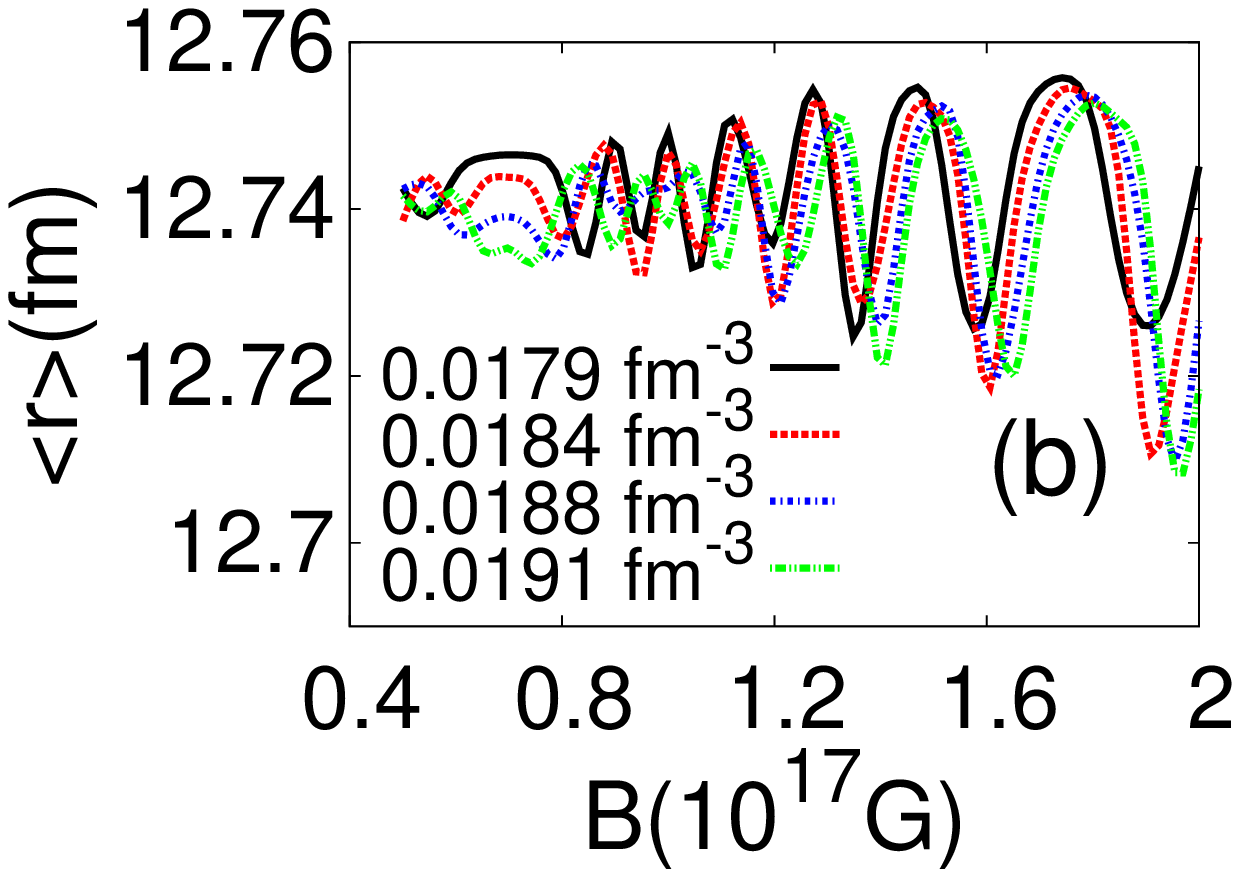} \\
\end{tabular}
	\caption{(Color online) Average radius for the droplet geometry, NL3 parametrization
       and proton fraction   $Y_{p}=0.3$:  (a) neutrons;
(b) electrons.
}\label{RaioMedioBarionGota}
\end{figure}

\begin{figure}[htb]\centering
\begin{tabular}{cc}
\includegraphics[width=0.48\linewidth,height=0.35\linewidth,angle=0]
{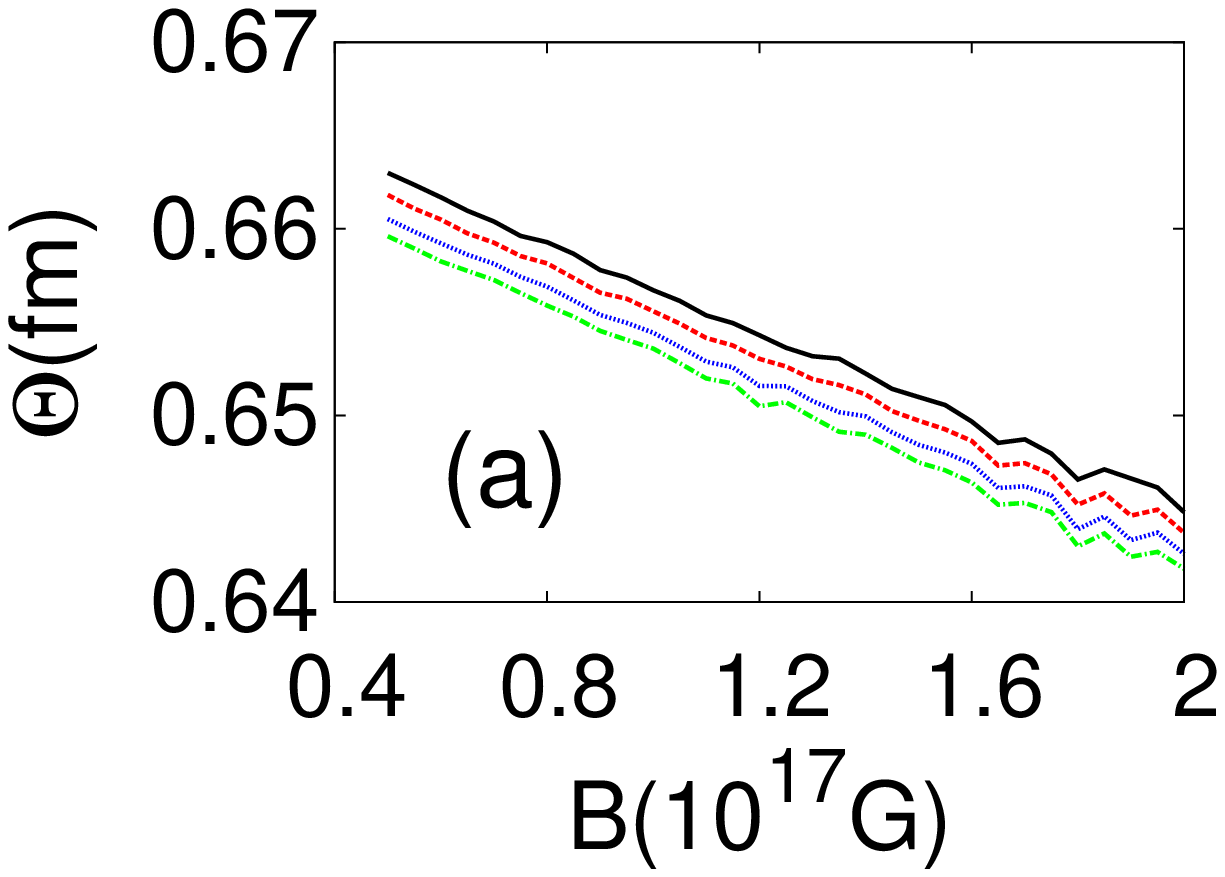}&
\includegraphics[width=0.5\linewidth,angle=0]
{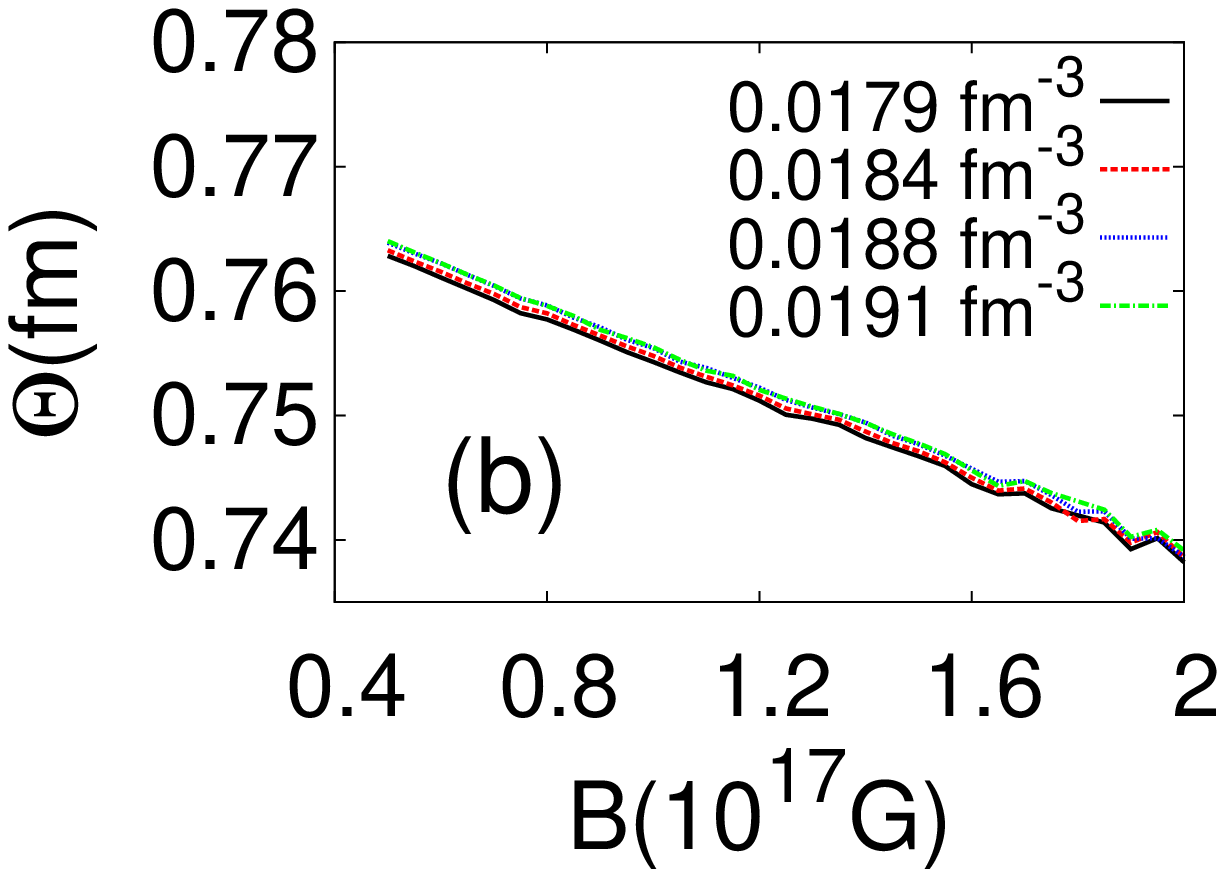}
\end{tabular}
\caption{(Color online) Neutron skin thickness for  NL3 and $Y_{p}=0.3$ for densities
  close to the drop-rod transition: (a) droplet geometry; (b) rod
  geometry.
}
\label{skin}
\end{figure}

In Fig. \ref{rhot} the transition densities between the geometries droplet-rod
(left), tube-bubble (right) are
plotted as a function of the field intensity. The transition density
between different geometries suffers  fluctuations that can be as high
as 5\% when the field changes between $10^{17}$~G and  $10^{18}$~G. However, taking fields not larger than $2\times 10^{17}$ the effect
on the transition between geometries is a reduction of the transition
density not be larger than 1.5\%.

\begin{figure}[htb]\centering
\begin{tabular}{ccc}
\includegraphics[width=0.45\linewidth,angle=0]  {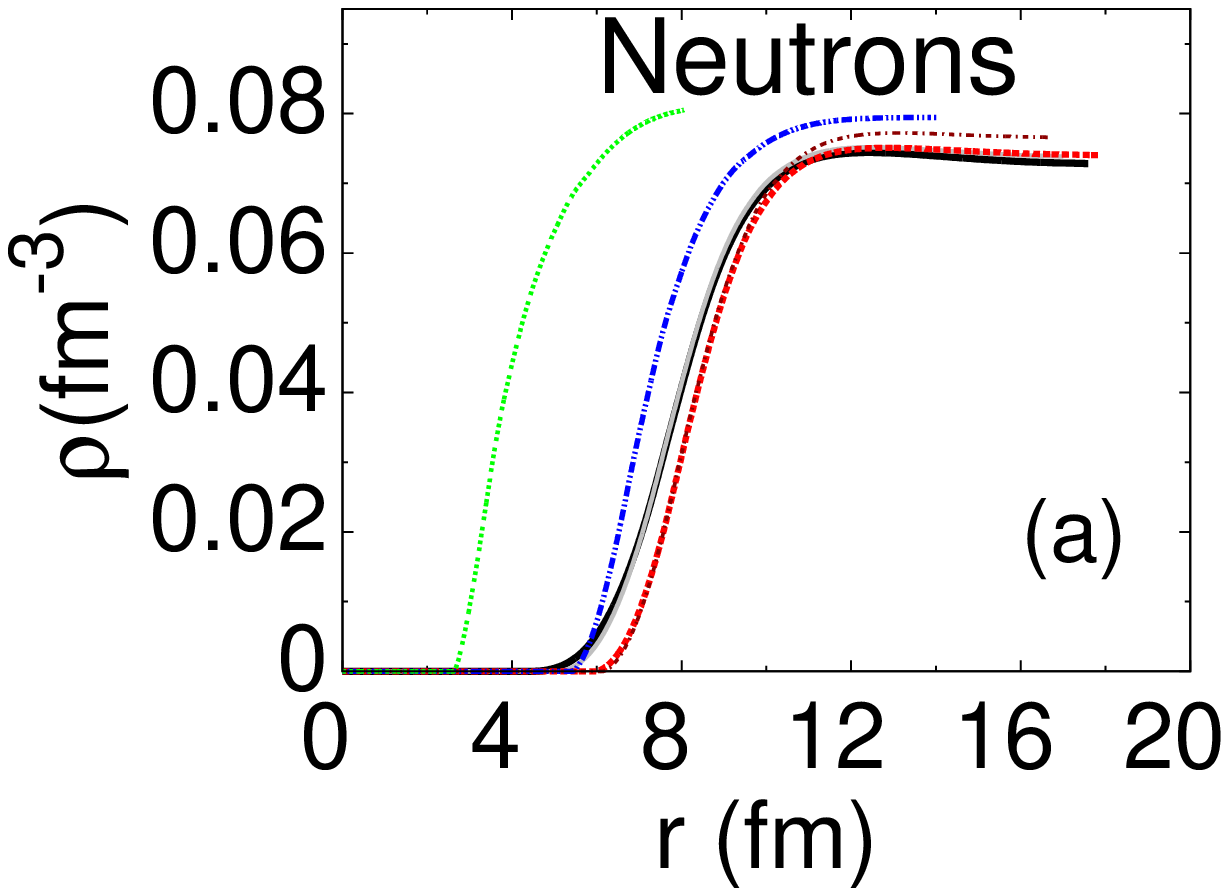} &
\includegraphics[width=0.45\linewidth,angle=0]  {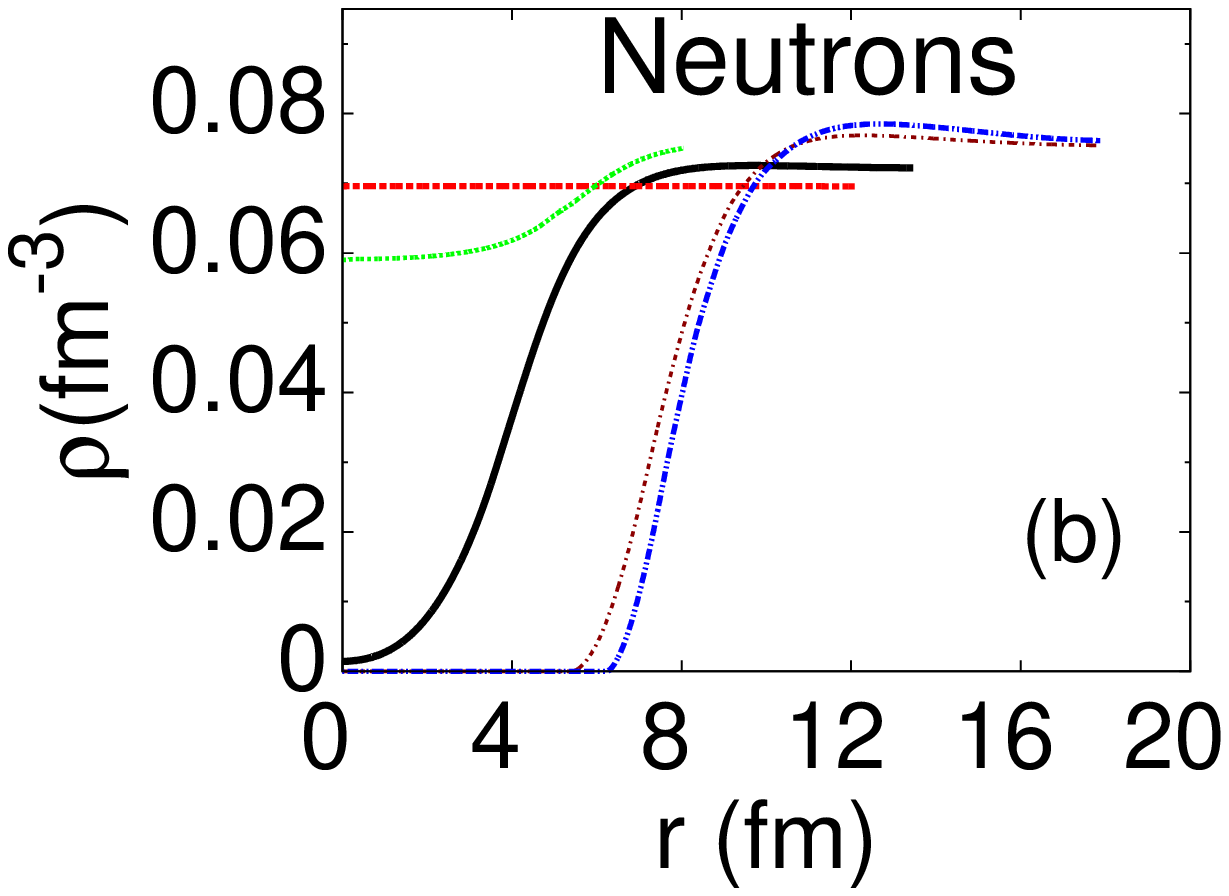} \\
\includegraphics[width=0.45\linewidth,angle=0]  {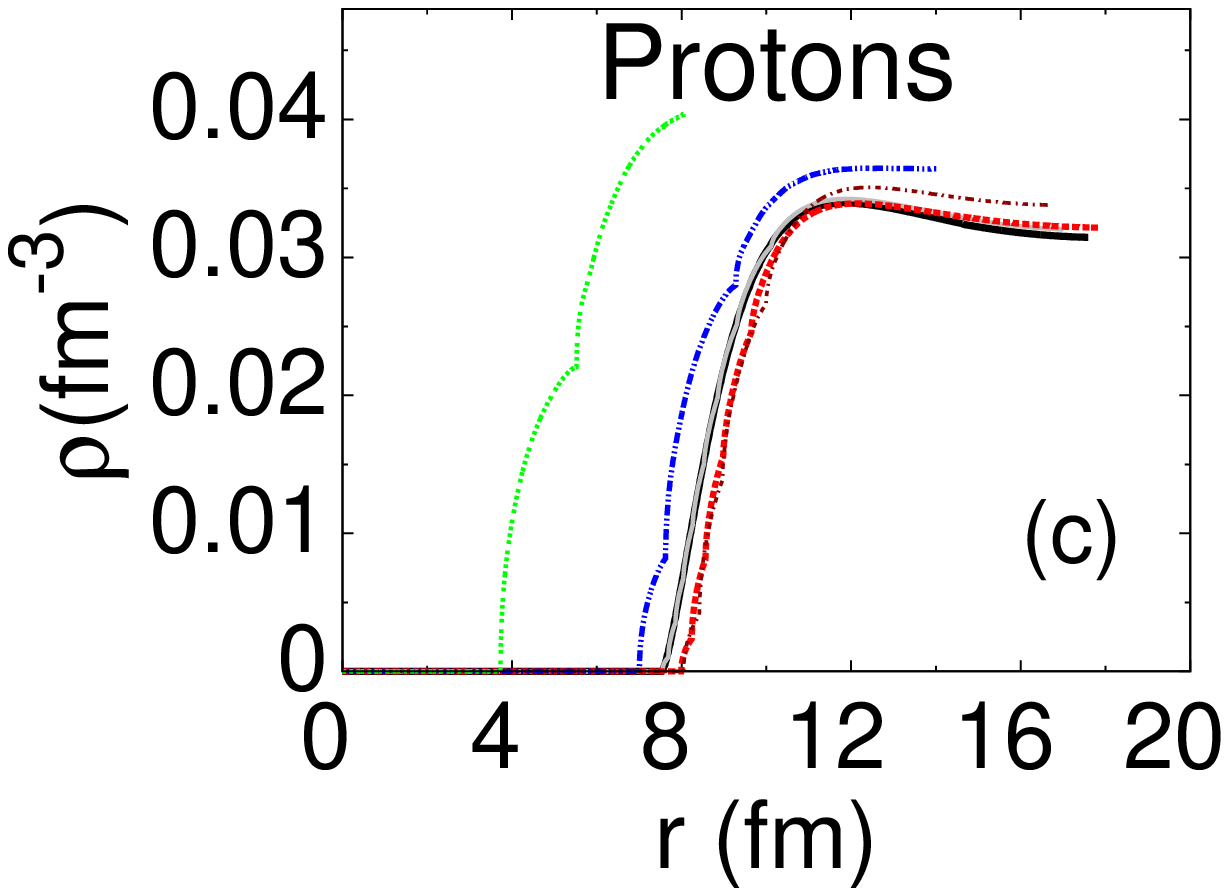} &
\includegraphics[width=0.45\linewidth,angle=0]  {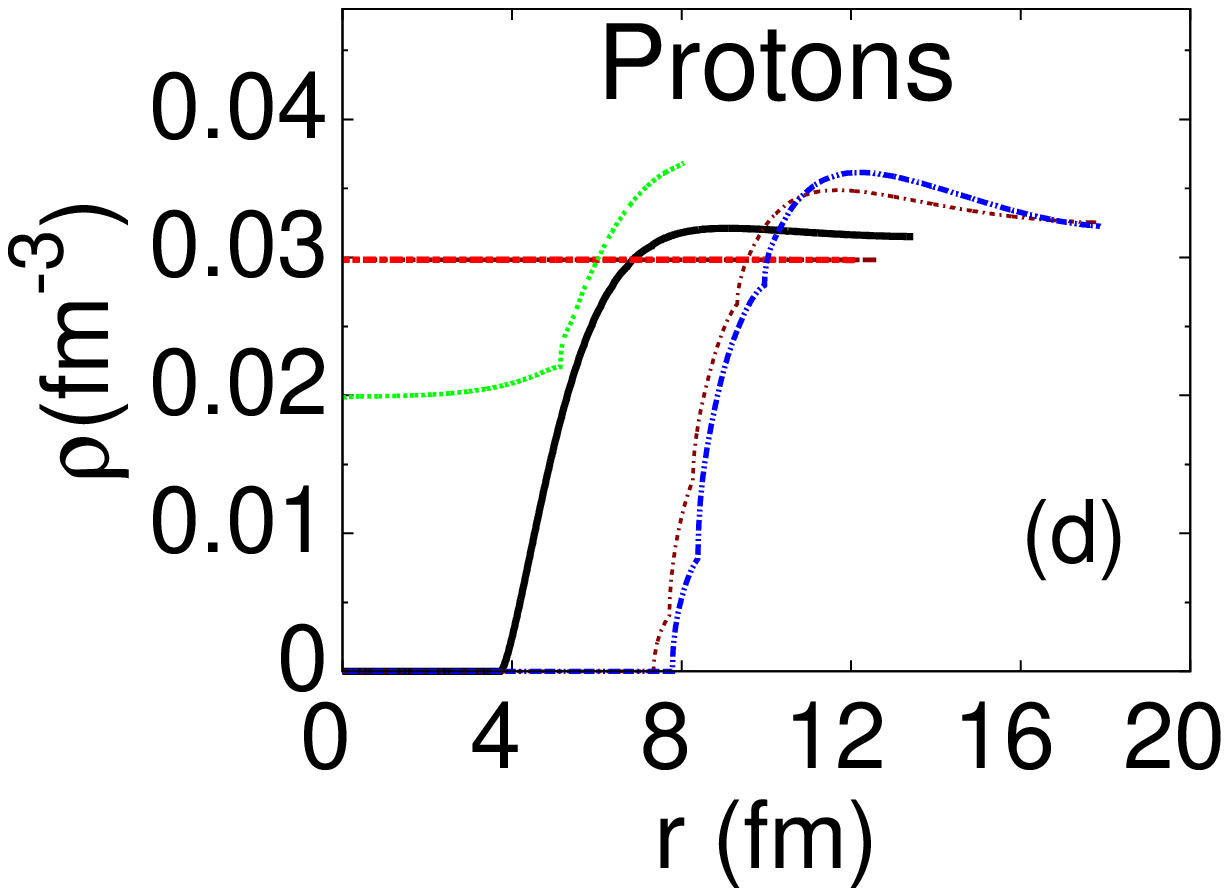} \\
\includegraphics[width=0.45\linewidth,angle=0]  {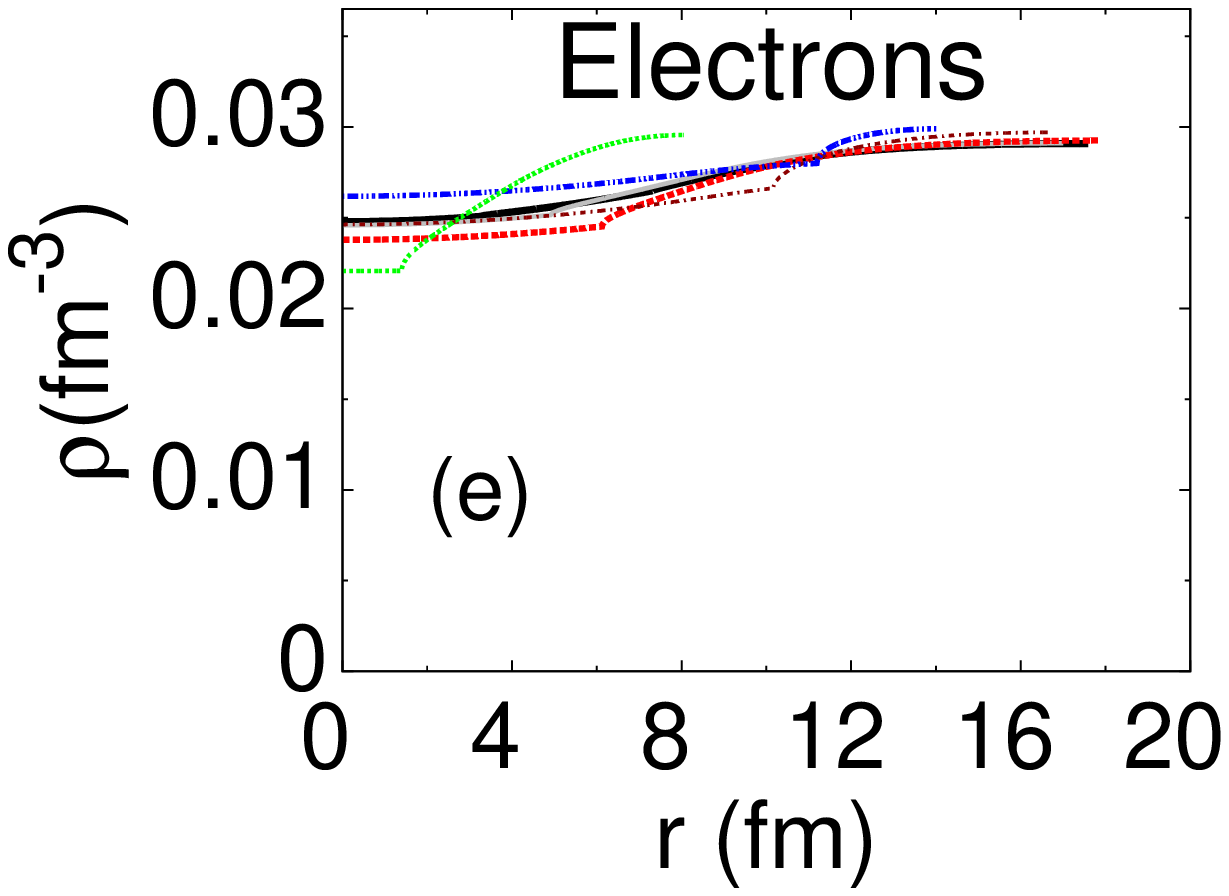} &
\includegraphics[width=0.45\linewidth,angle=0]  {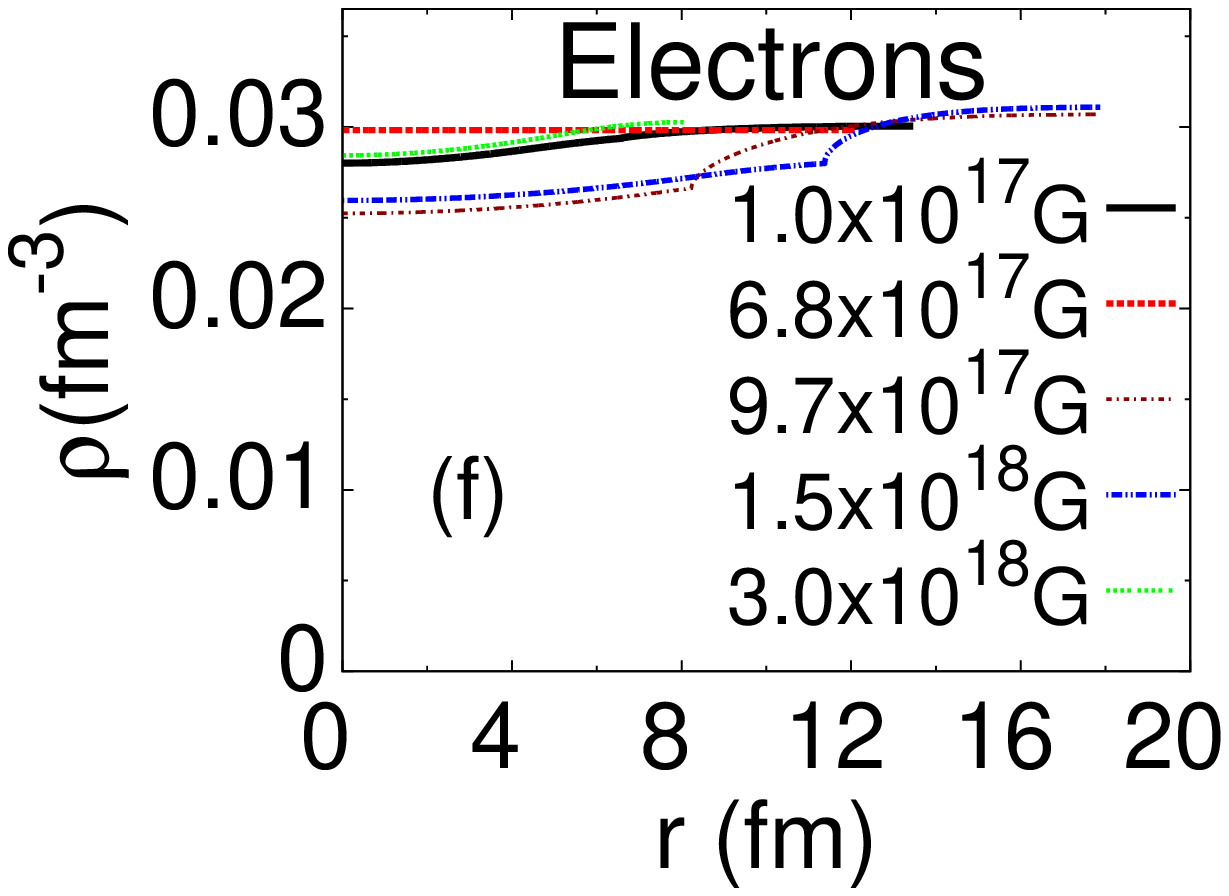} \\
\end{tabular}
	\caption{(Color online) Density profile in the Wigner-Seitz cell for the
          bubble geometry, NL3 parametrization, proton fraction
          $Y_{p}=0.3$ and baryonic density $\rho=0.095$ fm$^{-3}$ (left)
          and  $\rho=0.0995$ fm$^{-3}$ (right): (a,b) neutrons; (c,d)
       protons; (e,f) electrons.
}\label{DensidadeBubble}
\end{figure}

An increase of the binding energy between the nucleons gives rise
to a larger surface energy, which may  affect the pasta
structure namely the radius of the clusters, the crust-core transition   and 
 the transition between different
configurations. Moreover, Landau quantization may also gives origin to
large fluctuations. In Fig. \ref{sigma} 
the surface energy 
plotted for
parametrizations NL3 and TM1 and a proton fraction 
$Y_p=0.3$ for a slab configuration at $\rho=0.06$ fm$^{-3}$
 as a
function of the magnetic field intensity. The main trend is an increase of the
surface energy with the magnetic field. For  a field of the order of 10$^{18}$ G, which probably is
 already too
strong in the inner crust, the surface energy is 20\% larger when
compared with the no field case. A $2\times
10^{17}$ G field gives rise to an effect 10 times smaller, of the order of $\sim
2\%$.  These conclusions are confirmed by the top right panel of
Fig. \ref{free} where  the surface energy is
plotted as a function of the density at the drop-rod transition.

In Fig. \ref{ARwzSigma} results
 for the bubble configuration, including the
Coulomb interaction self-consistently, are plotted including  the surface
energy,  and the radius of  the Wigner Seitz cell. These
results were obtained for NL3 with the proton fraction $Y_p=0.3$ at $\rho=0.095$ fm$^{-3}$.  As already discussed
before, there is a clear increase of the surface energy with the magnetic
field intensity which can be as large as 50\% for $B=3\times 10^{18}$ G. Small
fluctuations of the surface energy reflect themselves on the Wigner Seitz cell
radius and total nucleon number inside the cluster. These effects are more
dramatic above $B=10^{18}$~G, but at $B\sim 2\times 10^{17}$G effects of
5-10\% are already expected.

In Fig. \ref{sigma2} the surface energy is plotted for densities below 
the drop-rod transition, and two proton fractions $Y_p=0.3$ and 0.1. For the
larger proton fraction we get a behavior similar to the one  previously
discussed, namely an increase of the surface energy  with the  increase of the
field strength. The surface energy is larger for the smaller densities because
clusters are smaller and the neutron dripped gas
is smaller. Decreasing the proton fraction to $Y_p=0.1$ this last feature is
still present (see Fig. \ref{yp01}), however, the surface energy suffers a small decrease for fields
below 10$^{18}$G and only increases for stronger fields.

In Fig. \ref{DensidadeDrop0.3} are shown  the density profiles of the configurations used
to calculate the surface energy for both proton fractions: for $Y_p=0.3$ the
thickness of the droplet surface decreases with $B$, while for the  $Y_p=0.1$
there is an increase of the surface thickness from 10$^{17}$ G to $5\times
10^{17}$G  followed by a decrease for still larger
fields. The number of Landau levels filled with the proton and electron
distributions for each field are given in Fig. \ref{LandauDrop0.1}. A smaller
number of levels is involved for the smaller proton fraction, and, therefore,
 $Y_p=0.1$ is more sensitive to strong magnetic fields.
The calculation of the surface energy reflects the size of the cluster, its proton fraction
and the interaction between particles. For $Y_p=0.3$ the size of the
cluster is pratically not affected  as seen in Fig. \ref{DensidadeDrop0.3}, and, therefore, $\sigma$ will
essencially give information about the binding between particles. On
the other hand, Fig. \ref{DensidadeDrop0.3} shows that if $Y_p=0.1$
the neutron distribution is quite affected by the interaction change
the protons feel in the presence of the magnetic field. On the whole
for weaker fields the surface energy decreases with $B$. For the larger magnetic  fields no neutrons drip out and the proton fraction of the droplet becomes 0.1. This very small fraction of protons favors smaller droplets, because a large asymmetry term reduces the stability of the clusters.

 In fact, increasing the magnetic field intensity changes the
  structure of the droplet pasta phase for a proton
  fraction $Y_p=0.1$ eliminating completly  the neutron dripped gas
  (see Fig. \ref{yp01}) and making the clusters less proton rich. A direct
  consequence is the transition to the rod geometry driven by  magnetic field.

In Fig. \ref{RaioMedioBarionGota} the average radius of the distribution of
neutrons and electrons inside a spherical  cluster is plotted for a set of 
densities close to the droplet-rod transition calculated according to
eq. (\ref{calcRaioMedio}). The main effect
of the magnetic field shows itself on the neutron distribution with an average
radius that decreases with $B$. Due to an
increase of the surface energy neutrons do not drip so easily and, therefore,
the number of neutrons outside the cluster is smaller.
Electrons are particularly sensitive to magnetic fields as strong as
10$^{17}$-10$^{18}$ G due to their small mass. The filling of Landau levels
gives rise to the fluctuations  shown on the right panel of
Fig. \ref{RaioMedioBarionGota}. This is a manifestation of the De Haas-van Alphen effect.

It is seen that the distribution of electrons is not flat and a 
self-consistent calculation that takes into account correctly charge
distribution will be affected by the magnetic field.
In particular, the rearrangement of the proton distributions will give rise to smaller proton
fractions at the cluster center and smaller neutron-skins. The effect
on the neutron-skins is seen in Fig. \ref{skin}  where the
neutron skin thickness calculated with NL3 for $Y_{p}=0.3$ and densities
  close to the drop-rod transition, according to eq. (\ref{nskin}), is plotted.  There is a decrease of about 3-4\% when the field increases from $5\times 10^{16}$ to  $2\times 10^{17}$ G. Above $B=10^{17}$ G the  oscillations present are a    consequence of the Landau quantization of the proton energy levels.

\begin{figure}[htb]\centering
\begin{tabular}{c}
\includegraphics[width=0.8\linewidth,angle=0]{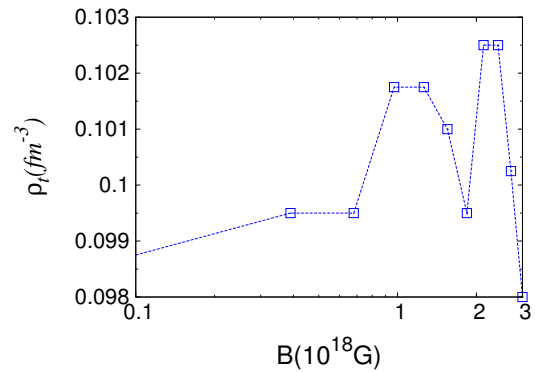}
\end{tabular}
	\caption{(Color online) Transition density at the bubble-homogeneous matter,
          for NL3  $Y_{p}=0.3$.}\label{rhot2}
\end{figure}

We next analyse the effect of the magnetic field on the crust-core transition.

The evolution of the Wigner Seitz radius and particle distributions inside the
cell with the magnetic field is 
plotted in  Fig. \ref{DensidadeBubble}  for two bubble configurations, at
$\rho=0.095$ and 0.0995 fm$^{-3}$, the second one
close to the crust core transition. For $10^{17}<B<10^{18}$ G and $\rho=0.095$
fm$^{-3}$, the variation of $R_{WS}$ is small and becomes
more pronounced for stronger fields. However, it is clearly seen the
effect of B on the surface energy which gives rise to a narrower surface thickness, larger
central densities, and smaller tails at the bubble center. This is true for
both neutrons and protons. As
expected due to their much smaller mass, the electron distributions  are more sensitive to the field
intensity. 

The density $\rho=0.0995$ fm$^{-3}$ is very close to the crust-core
transition, and the filling of the Landau levels may dictate that for a given
$B$ the transition to the core has already occured ($B=6.8\times 10^{17}$ G) while for other intensities
a more pronounced bubble occurs with smaller electron densities in the center
of the bubble ( $B\sim 10^{18}$ G). For $B=3\times 10^{18}$ G a
configuration very close to the crust-core transition occurs.
 The
transition density to the homogeneous phase suffers essentially an
increase that will be of  1\% for  $2\times 10^{17}$ G and can go up to
5\% for  $B=10^{18}$ G, see Fig. \ref{rhot2}. 

Magnetic field fluctuations may give rise to non-equilibrium configurations
that will tend to evolve in time to equilibrium configurations  originating
inner tensions that could give rise to star-quakes and bursting activity of magnetars.

\section{Conclusion} \label{conclusion}
%
%
In the present study the effect of the magnetic field on the pasta phase
calculated within a Thomas Fermi formalism has been discussed. Nuclear matter
was described mostly by the RMF parametrization NL3 and  proton fractions of
0.3 and 0.1 were considered. 

Our main aim was to determine how the magnetic field could affect the free
energy per particle, the radius of the Wigner Seitz cells, the cluster
properties, and the transitions between different configurations, or the
crust-core transition. Most of the calculations were done for fields below
10$^{18}$ G,  although, in order to estimate upper limits, some of
the calculations were pushed to $3\times 10^{18}$ G.

It is known that the pasta phase is a frustrated system that results from the
competition between the Coulomb  and the surface energy. It is, therefore,
expected that this phase will be affected by a strong magnetic field. Charged
particles in a 
magnetic field suffer the Landau quantization which gives rise to a decrease
of the free energy per particle due to the large degeneracy levels in the
direction perpendicular to the field, and, therefore, to an increase of the
surface energy. The surface thickness of clusters will be thinner, the inner
densities larger and, since neutrons will not drip off so easily, a smaller
number of particles will occur in the background gas. We have also shown that
the transition between different configurations or crust-core will  be
affected although in an irregular way.  Fluctuations of the magnetic field may
give rise to inner stresses that oblige the system to evolve to an equilibrium
configuration and originate  bursting  activity of magnetars.

In the present work was just exploratory and a more careful study should be
done that uses models with a symmetry energy that satisfies experimental
constraints and considers smaller proton fraction, namely $\beta$-equilibrium
stellar matter. A study of the stress developed on these structures
should also be performed.

\section*{Acknowledgments}
This work was partially supported by COMPETE/FEDER and FCT (Portugal) under the grant
PTDC/FIS/113292/2009, CNPQ (Brazil) and Capes/FCT (Brazil) under project 232/09 and FAPESC (Brazil) under the grant 6316/2011-9.


\begin{thebibliography}{29}
\bibitem{pethick}     D. G. Ravenhall, C. J. Pethick, and J. R. Wilson, Phys. Rev. Lett. 50, 2066 (1983).
\bibitem{hashimoto84}     M. Hashimoto, H. Seki, and M. Yamada, Prog. Theor. Phys. 71, 320 (1984).
 \bibitem{hor05}    C. J. Horowitz, M. A. P\'{e}rez-Garcia, and J. Piekarewicz, Phys. Rev. C 69, 045804 (2004); C. J. Horowitz, M. A. P\'{e}rez-Garcia, D. K. Berry, and J. Piekarewicz, ibid. 72, 035801 (2005).
\bibitem{maruyama05}     T. Maruyama, T. Tatsumi, D. N. Voskresensky, T. Tanigawa, and S. Chiba, Phys. Rev. C 72, 015802 (2005).
\bibitem{watanabe08}     G. Watanabe, K. Sato, K. Yasuoka, and T. Ebisuzaki, Phys. Rev. C 66, 012801 (2002); 68, 035806 (2003); 69, 055805 (2004); H. Sonoda, G. Watanabe, K. Sato, K. Yasuoka, and T. Ebisuzaki, ibid. 77, 035806 (2008).
\bibitem{grill2012} F. Grill, C. Provid\^{e}ncia, and S. S. Avancini
Phys. Rev. C 85, 055808 (2012).
\bibitem{bb}  J. Boguta and A. R. Bodmer, Nucl. Phys. A  292, 413 (1997).

\bibitem{pasta1}     S. S. Avancini, D. P. Menezes, M. D. Alloy, J. R. Marinelli, M. M. W. Moraes, and C. Provid\^{e}ncia, Phys. Rev. C 78, 015802 (2008).
\bibitem{pasta2}    S. S. Avancini, L. Brito, J. R. Marinelli,
  D. P. Menezes, M. M. W. de Moraes, C. Provid\^{e}ncia, and
  A. M. Santos, Phys. Rev. C 79, 035804 (2009).
\bibitem{menezes99} D. P. Menezes and C. Provid\^{e}ncia, Phys. Rev. C 60,
  024313 (1999)

\bibitem{pasta3} S. S. Avancini, S. Chiacchiera, D. P. Menezes, and C. Provid\^{e}ncia
Phys. Rev. C 82, 055807 (2010); Erratum Phys. Rev. C 85, 059904(E).
\bibitem{duncan}R. C. Duncan and C. Thompson, Astrophys. J. 392, L9 (1992); C. Thompson and R. C. Duncan, MNRAS 275, 255 (1995). 
\bibitem{usov}V. V. Usov, Nature (London) 357, 472 (1992).
\bibitem{pacz}B. Paczy\'{n}ski, Acta Astron. 42, 145 (1992).
\bibitem{index} McGill SGR/AXP Online Catalog, [http://www.physics.mcgill.ca/~pulsar/magnetar/main.html]. 
\bibitem{virial} D. Lai and S. Shapiro, ApJ, 383, 745 (1991).
\bibitem{broderick02}A. E. Broderick,M. Prakash, J. M. Lattimer, Phys. Lett. B 531, 167 (2002)
\bibitem{sedrakian13} M. Sinha, B. Mukhopadhyay, A. Sedrakian,  Nucl. Phys. A 898, 43 (2013).
\bibitem{chakrabarty97} D. Bandyopadhyay, S. Chakrabarty, and S. Pal, Phys. Rev. Lett. 79, 2176 (1997).
\bibitem{nl3}    G. A. Lalazissis, J. K\"{o}nig, and P. Ring, Phys. Rev. C 55, 540 (1997).
\bibitem{tm1}    K. Sumiyoshi, H. Kuwabara, and H. Toki, Nucl. Phys. A 581, 725 (1995).
\bibitem{Sugahara1994}  Y. Sugahara, and H. Toki, Nucl. Phys. A 579, 557 (1994).
\bibitem{aziz08} A. Rabhi, C. Provid\^{e}ncia and J. da Provid\^{e}ncia, J. Phys. G35, 125201 (2008) doi:10.1088/0954-3899/35/12/125201
\bibitem{broderick00} A. Broderick, M. Prakash, and J. M. Lattimer, Astrophys. J. 537, 351 (2000).
\bibitem {landau} L. D. Landau and E. M. Lifshitz, Quantum Mechanics, Volume 3 of A Course of Theoretical Physics, Pergamon Press, 1965. 
\bibitem{walecka} H. M\"{u}ller and B. D. Serot, Phys. Rev. C 52, 2072 (1995). 
\bibitem{aziz09} A. Rabhi, C. Provid\^{e}ncia, and J. da Provid\^{e}ncia, Phys. Rev. C 79, 015804 (2009); Aziz Rabhi and C.  Provid\^{e}ncia  Journal of Physics G: Nuclear and Particle Physics  37,
075102 (2010).
\bibitem{pais2012} H. Pais and J. R. Stone, Phys. Rev. Lett. 109, 151101 (2012) 
\bibitem{dorso2012} C. O. Dorso, P. A. G. Molinelli, and J. A. L\'{o}pez, Phys. Rev. C 86, 055805 (2012).




\end{thebibliography}
\end{document}